\documentclass[a4paper,11pt]{article}

\usepackage{graphicx} 

\pdfoutput=1 

\usepackage{jcappub}
\usepackage{bbold}
\usepackage{subcaption}
\usepackage{float}
\usepackage{appendix}
\usepackage{multirow}
\usepackage{amsmath}
\usepackage{placeins}
\usepackage{orcidlink}
\usepackage[normalem]{ulem}

\usepackage{titlesec} 
\titlespacing*{\subsubsection}{0pt}{1em}{2em}
\newcommand{\be}{\begin{equation}}
\newcommand{\ee}{\end{equation}}
\definecolor{DarkGreen}{rgb}{0,0.6,0}
\definecolor{SkyBlue}{rgb}{0, 0.5, 0.8}
\definecolor{gray}{rgb}{0.5, 0.5, 0.5}

\usepackage[capitalise]{cleveref}

\title{Model-independent consistency tests of DESI DR2 BAO and SN Ia}

\author[a,b]{Hyeok Woo\,\orcidlink{0009-0009-1510-9379}}
\author[a]{William Luke Matthewson\,\orcidlink{0000-0001-6957-772X}}
\author[a,b]{Arman Shafieloo\,\orcidlink{0000-0001-6815-0337}}

\affiliation[a]{Korea Astronomy and Space Science Institute, 776, Daedeokdae-ro, Yuseong-gu, Daejeon 34055, Republic of Korea}
\affiliation[b]{University of Science and Technology, 217 Gajeong-ro, Yuseong-gu, Daejeon 34113, Republic of Korea}

\emailAdd{wooh@kasi.re.kr}
\emailAdd{willmatt4th@kasi.re.kr}
\emailAdd{shafieloo@kasi.re.kr}

\date{July 2025}

\begin{document}

\abstract{
Cosmic distances can be measured using two complementary probes: Type Ia supernovae (SN Ia), serving as standard candles, and baryon acoustic oscillations (BAO), serving as standard rulers. The luminosity distance derived from supernovae and the angular diameter distance obtained from BAO must be mutually consistent if these data are to be combined for cosmological inference. Hence, the existence of potential discrepancies, whether arising from systematics in either dataset or from violation of the cosmic duality relation (in an unconventional cosmology), remains an important issue to address. Testing consistency under a particular cosmological model can be limiting, as the model may not be sensitive to every kind of inconsistency possible in the data. Thus, in this work we use a model-independent Crossing Statistics framework to test the consistency, using DESI DR2 BAO, and the Pantheon+ and Union3 SN Ia datasets. We find adding up to two additional degrees of freedom, using Crossing Statistics on the $\Lambda$CDM distance-redshift relation, to be statistically justified. In these cases, the two probes remain mutually consistent at the $1\sigma-2\sigma$ level. Having established this statistical consistency, we combine the datasets to reconstruct the expansion history of the Universe and the inferred evolution of dark energy. 
The reconstructions obtained using different crossing variables show compatible behaviour where the data constraints are strongest, particularly at low redshift. 
Overall, the results are suggestive of a dark energy component that is evolving at low redshift, compatible with results from other reconstructions methods.}

\maketitle
\section{Introduction}

The accelerated expansion of the Universe has been established through various cosmological probes, most notably Type Ia supernovae (SN Ia) as standard candles and baryon acoustic oscillations (BAO) as standard rulers~\cite{SupernovaSearchTeam:1998fmf, Perlmutter:1999jt, previous_BAO_measurement}. These two datasets provide complementary constraints on the expansion history and play a crucial role in supporting the current concordance cosmological model, $\Lambda$CDM.
The luminosity distance $D_{L}(z)$ from supernovae and the angular diameter distance $D_{A}(z)$ from BAO must satisfy the cosmic distance duality (or Etherington) relation $D_{L}(z)=(1+z)^{2}D_{A}(z)$ for metric theories of gravity~\cite{2007GReGr..39.1047E,Ellis:2007}. Therefore, measurements of BAO and SN Ia must remain mutually consistent when mapped into the same observable space. Any statistically significant deviation could be indicative of systematic uncertainties or new physics, that are not accounted for.

Most consistency studies rely on a specific cosmological model, typically $\Lambda$CDM or simple extensions of it. However, such approaches can obscure genuine discrepancies; the model itself might restrict the behaviours that can be probed in the data, and thus produce seemingly consistent results, even when there are inconsistencies in the raw data. A more robust strategy is to use model-independent tests that directly compare the cosmological information encoded in the data, in our case the distance-redshift relation. 

Several recent works have used such approaches in the context of the DESI DR2 BAO data and existing supernovae compilations. For example, Gaussian Process (GP) reconstructions of the expansion history have been used to perform tests of consistency between late-time probes. GP regression can reconstruct the relevant functions independently of a cosmological model. In particular, the recent analysis in \cite{Dinda:2025hiu, Dinda:2026uff} makes use of an Alcock-Paczynski variable, which also allows BAO and SN Ia measurements to be compared without assuming a calibration for the sound horizon or supernova absolute magnitude. Their latest analysis found that current SN Ia and BAO datasets are broadly consistent with each other across their redshift ranges~\cite{Dinda:2026uff}. While GP methods provide flexible, non-parametric reconstructions, they can depend on the choice of kernel and mean function, as well as the marginalisation over the hyperparameters, which can influence the inferred behaviour, particularly in sparsely sampled redshift regions~\cite{2012PhRvD..85l3530S, Hwang:2022hla}. It is also not trivial to perform model comparison, as the effective number of degrees of freedom of the GP reconstruction is not well-known.
 
The Bayesian implementation of Crossing Statistics~\cite{2012JCAP...05..024S}, provides an effective framework for this purpose, by introducing particular deformation freedoms to a chosen fiducial expansion history. These deformations are controlled by the coefficients of Chebyshev polynomial basis functions, have a well-defined number of additional degrees of freedom, and are always reducible to the (nested) fiducial case. By testing whether independent datasets prefer compatible deformation patterns in the extended parameter space, we can obtain a complementary cross-check of mutual consistency between SN Ia and BAO data.

The recent analyses within DESI of the DR2 BAO data have also reported indications that the inferred expansion history may prefer departures from a cosmological constant description, whether interpreted within a specific dark energy parametrization, or using model-independent methods~\cite{DESI:2025zgx, DESI:2025fii}. These developments highlight the importance of extracting the cosmological information encoded in different late-time probes without relying on a specific cosmological model~\cite{2012JCAP...05..024S}, as an apparent lack of discrepancies may depend strongly on the sensitivity of the assumed model to different behaviours~\cite{2008IJMPD..17.2315L,2010PhRvD..82j3502H}.

Another recent work, focusing on reconstruction, tries to determine the evolution of the dark energy equation of state using Weighted Function Regression applied to a combination of DESI DR2 BAO, supernova and Planck cosmic microwave background (CMB) datasets~\cite{Gonzalez-Fuentes:2025lei}. This model-independent analysis reports a $2\sigma$ - $4\sigma$ preference for a dark energy evolution that crosses $w=-1$, in agreement with the results from the DESI analysis. 

Crossing Statistics may also be used to investigate the behaviour of dark energy, through model-independent reconstructions of the equation of state, or relative energy density~\cite{DESI:2024aqx}. Indeed, a related Crossing Statistics analysis was performed in DESI, where the method was applied to reconstruct the potential evolution of the relative dark energy density from the most recent BAO data release~\cite{DESI:2025fii}. However, the above works focus on reconstruction of dark energy behaviour without first testing consistency between independent probes, and do not implement the reconstruction directly in the observable $D_L(z)$ space, favouring instead the equation of state parameter, or relative dark energy density. 

In this work, we investigate the consistency between SN Ia and BAO datasets in a model-independent manner, using Crossing Statistics. Both datasets are mapped to the luminosity-distance space to enable a direct comparison, and the deformation parameters are restricted to preserve the present-day Hubble parameter and defined with respect to the various data redshift ranges. We systematically increase the number of free parameters in the deformations, to see at what level the resulting, more flexible behaviours agree across the datasets.
We utilize the Pantheon+ and Union3 SN Ia samples, together with the Dark energy Spectroscopic Instrument (DESI) DR2 BAO measurements, covering a broad redshift range.

This paper is set out as follows. In \cref{s:meth}, we introduce the methodology and data used in the analysis. Then, in \cref{s:res}, we display our results. Finally, we discuss our findings and summarise our conclusions in \cref{s:conc}.

\section{Data and Methodology}
\label{s:meth}
\subsection{Background cosmology and observables}

We model the late-time Universe as a spatially homogeneous and isotropic Friedmann-Lemaître-Robertson-Walker (FLRW) spacetime within general relativity, containing effective components such as non-relativistic matter and dark energy. Throughout this work we assume spatial flatness, $k=0$, so that the expansion history is governed by the Friedmann equation

\begin{equation}
H^2(z)
=
H_0^2 \left[
\Omega_{m,0}(1+z)^3
+
\Omega_{r,0}(1+z)^4
+
\Omega_{{\rm DE},0}\, \frac{\rho_{\rm DE}(z)}{\rho_{{\rm DE},0}}
\right]\,,
\label{eq:friedmann_fde}
\end{equation}

where $H_0$ is the present-day value of the Hubble parameter and 
$\Omega_{m,0}$, $\Omega_{r,0}$, and $\Omega_{{\rm DE},0}$ 
denote the present-day density parameters of matter, radiation, 
and dark energy, respectively, satisfying
$\Omega_{m,0}+\Omega_{r,0}+\Omega_{{\rm DE},0}=1$.
$\rho_{\rm DE}(z)$ denotes the physical dark energy density, 
and $\rho_{{\rm DE},0}$ its present-day value.
Here we introduce the dimensionless relative dark energy density, $f_{\rm DE}(z)$, defined such that
\begin{equation}
f_{\rm DE}(z) \equiv 
\frac{\rho_{\rm DE}(z)}{\rho_{{\rm DE},0}}=
\exp\!\left[3\int_{0}^{z}\frac{1+w(z')}{1+z'}\,dz'\right]\,,
\end{equation}
where $w(z)$ is the dark energy equation of state.
In the $\Lambda$CDM model, the dark energy equation of state is constant, $w=-1$, which leads to a redshift-independent dark energy density, so $f_{\rm DE}=1$.
 More generally, phenomenological extensions allow for a time-varying equation of state. In particular, the commonly-used $w_0w_a$ parametrization~\cite{Chevallier:2000qy,Linder:2002et}, is given by
\begin{equation}
w(z)=w_0 + w_a \frac{z}{1+z}\,.
\end{equation}
While such parametrizations provide a convenient framework for exploring departures from a cosmological constant, they impose specific functional choices, which may limit their sensitivity to certain behaviours. In contrast, the Crossing Statistics approach adopted in this work does not rely on any single form of $w(z)$, but instead can probe deviations directly at the level of the luminosity distance-redshift relation, up to arbitrarily high order in principle.

Cosmological distance measures depend on the background expansion history and are probed here using two complementary observables: Type~Ia supernovae (SN Ia), which act as standard candles, and baryon acoustic oscillations (BAO), which provide a standard ruler. SN Ia measurements are made  using the apparent magnitude $m_B$, which is related to the luminosity distance through the definition
\begin{equation}
D_L(z)=10^{(m_B-M_B+5)/5}\,\mathrm{pc}\,,
\label{eq:pogson}
\end{equation}
where $M_B$ is the absolute magnitude, treated as a nuisance parameter.

BAO measurements constrain the comoving transverse distance $D_M(z)$, the Hubble distance
$D_H(z)\equiv c/H(z)$, and the combined volume-averaged distance
\begin{equation}
D_V(z)\equiv \left[z\,D_M^2(z)\,D_H(z)\right]^{1/3}\,,
\end{equation}
which are measured as dimensionless ratios $D_M/r_d$,
$D_H/r_d$, and $D_V/r_d$, where $r_d$ denotes the sound horizon at the drag epoch.

In a spatially flat universe, the transverse comoving distance is given by
\begin{equation}
D_M(z)=c\int_{0}^{z}\frac{dz'}{H(z')}\,.
\label{eq:DM_def}
\end{equation} 

To place BAO and SN Ia on a common footing, we map the BAO distance constraints into the luminosity-distance space using the cosmic distance duality relation
\begin{equation}
D_L(z)=(1+z)\,D_M(z)=(1+z)c\int_{0}^{z}\frac{dz'}{H(z')}\,,
\label{eq:DL_def}
\end{equation}
which holds in any metric theory of gravity under photon number conservation.
This procedure enables a direct comparison of the two datasets with the fewest assumptions possible about the underlying cosmological model. Any deviation between SN Ia and BAO can hence be compared using the same quantity, $D_L(z)$.

In addition to the luminosity-distance relation and the relative dark energy density $f_{\rm DE}(z)$, we also examine derived kinematic diagnostics of the expansion history. In particular, we consider the $Om(z)$ diagnostic~\cite{Sahni:2008xx}
\begin{equation}
Om(z) \equiv \frac{H^2(z)/H_0^2 - 1}{(1+z)^3 - 1}\,,
\end{equation}
which provides a null test of $\Lambda$CDM. In a spatially flat universe with a cosmological constant, $Om(z)$ is constant and equal to $\Omega_{m,0}$. 
Therefore, deviations from this constant value indicate departures from a $\Lambda$CDM expansion history.
We also consider the deceleration parameter
\begin{equation}
q(z) \equiv -\,\frac{\ddot a a}{\dot a^2}
= \frac{d \ln H(z)}{d \ln (1+z)} - 1\,,
\end{equation}
which characterizes the acceleration of the Universe. 
A negative value of $q(z)$ corresponds to accelerated expansion, while positive values indicate that the expansion is decelerating. 
These complementary diagnostics help with the physical interpretation of possible deviations from $\Lambda$CDM in the reconstructed expansion history.

\subsection{Datasets}
\subsubsection{Type Ia supernovae}

We employ two state-of-the-art Type~Ia supernova (SN Ia) compilations, Pantheon+ and Union3, which together provide high-precision measurements of the luminosity distance relation over a wide redshift range.

\begin{itemize}
\item \textbf{Pantheon+}:
The Pantheon+ compilation~\cite{Brout:2022vxf} consists of spectroscopically confirmed SN Ia spanning the redshift range $0.01 \lesssim z \lesssim 2.26$. 
It represents a significant improvement over earlier Pantheon releases through updated photometric calibration, refined light-curve fitting, and a comprehensive treatment of systematic uncertainties.
Standard quality cuts and calibration selections are applied as described in the original Pantheon+ analysis, resulting in a final set of 1590 supernovae with $(z \geq 0.01)$.

\item \textbf{Union3}:
The Union3 compilation~\cite{2016AAS...22713918R} contains 2087 SN Ia, combining data from multiple surveys with consistent calibration and systematic corrections. 
Unlike Pantheon+, Union3 distances are derived using a Bayesian hierarchical framework, implemented through the UNITY formalism, in which the redshift dependence of the luminosity distance is summarized by a set of spline nodes ($22$ effective nodes) rather than individual supernovae.
Although Pantheon+ and Union3 share a large subset of underlying supernovae, their independent analysis pipelines and distance reconstruction methodologies make them complementary for consistency tests.
\end{itemize}

The \textbf{DES-Y5} (Dark Energy Survey Year 5) SN sample~\cite{DES:2024jxu} is excluded from the primary analysis. Its substantially lower maximum redshift compared to Pantheon+ and Union3 leads to a different normalization of the crossing hyperparameters, making a direct comparison within the full redshift range difficult, when using Crossing Statistics framework~\cite{Matthewson_2025}. To assess its consistency with BAO measurements, we perform a dedicated low-redshift analysis, in which the DESI DR2 dataset is restricted to a comparable redshift range; the results are presented in Appendix~\ref{a:DES}.

\subsubsection{Baryon acoustic oscillations}

\begin{itemize}
\item \textbf{DESI DR2}:
The baryon acoustic oscillations measurements are taken from the DESI Data Release~2 (DR2)~\cite{DESI:2025zgx}, and provide high-precision constraints on the comoving transverse distance $D_M(z)$ and the Hubble distance $D_H(z)=c/H(z)$ across multiple redshift bins.
The DESI DR2 BAO sample includes 13 effective redshift bins derived from different tracer populations, enabling distance measurements over the approximate range $0.3 \lesssim z \lesssim 2.33$. BAO observables are reported in the dimensionless combinations $D_M/r_d$, $D_H/r_d$, and, in one case, $D_V /r_d$. 
Since cosmic microwave background (CMB) data are not included in this analysis, we treat $r_d$ as a free nuisance parameter and marginalize over it.

\end{itemize}

\subsection{Crossing Statistics}

Consistency tests performed strictly within one model's framework, for example $\Lambda$CDM, may obscure discrepancies between SN Ia and BAO, because the assumed background model enforces a limited set of behaviours for the distance–redshift relation. If the datasets differ in a way that the model cannot capture, their constraints may nevertheless appear artificially consistent. This is equally true for $\Lambda$CDM and other extended models. To avoid this limitation, we employ the Bayesian application of Crossing Statistics~\cite{2012JCAP...05..024S,2013JCAP...04..042S}, in which the fiducial luminosity distance is modified by a smooth, model-agnostic deformation. This enables us to test whether the two probes favour the same underlying distance relation behaviour, without assuming a specific dark energy parametrization or modified-gravity model.

The key idea is that, if SN Ia and BAO encode mutually consistent cosmological information, they should prefer overlapping posterior regions in the extended parameter space of crossing parameters ${C_{n}}$ even after additional flexibility is introduced. Conversely, persistent differences in their preferred crossing coefficients would point to residual systematics or genuine physical inconsistencies~\cite{2013JCAP...04..042S,Matthewson_2025}. This method has been tested and used previously. The approach adopted here modifies it slightly, introducing crossing deformations at the level of the luminosity distance itself, and enabling a direct comparison of SN Ia and BAO observables within a common framework, while simultaneously assessing their mutual consistency.

\subsubsection{Deformation prescription}

We choose $\Lambda$CDM as our fiducial starting point, and deform the associated luminosity distance $D^{\Lambda{\rm CDM}}_{L}(z)$ using Chebyshev polynomials as follows~\cite{2012JCAP...05..024S}:

\be
D^{\times}_{L}(z) = D^{\Lambda{\rm CDM}}_{L}(z) \times \sum_{n=0}^{N} C_{n}T_{n}(\tilde{z})\,,
\ee
where $T_{n}(\tilde{z})$ are Chebyshev polynomials of the first kind,

\[
\begin{aligned}
T_{0}(x) &= 1\,, \\
T_{1}(x) &= x\,, \\
T_{n+1}(x) &= 2x T_{n}(x) - T_{n-1}(x)\,,\\
&{\rm etc.}
\end{aligned}
\]
and $\tilde{z}$  is the redshift, rescaled to the interval \([-1, 1]\):

\be
\tilde{z} \equiv 2\frac{z}{z_{\max}} - 1\,,
\label{eq:normalized_redshift}
\ee
where $z_{max}$ is chosen as the maximum redshift of the combined dataset. Normalizing each dataset to the same redshift interval ensures that the deformation coefficients $C_{n}$ retain a consistent interpretation across all probes.

A polynomial basis in this  prescription is useful in various ways.
First, the Chebyshev polynomials form an orthogonal set over the interval $[-1,1]$, which helps to minimize correlations among the deformation coefficients ${C_i}$. 

Second, the ordering of the Chebyshev modes provides a natural hierarchy of deformations: low-order modes correspond to smooth, global modifications of the luminosity distance relation, while higher-order modes capture progressively more localized features in redshift. 
This property allows for a controlled and systematic introduction of flexibility, making it possible to detect the onset of overfitting as additional degrees of freedom are included.

Importantly, this construction does not impose any specific physical model for deviations from $\Lambda$CDM, but instead provides a flexible, data-driven description of departures in the distance-redshift relation~\cite{Shafieloo:2012ht}.\\

\subsubsection{Additional constraints}
\label{s:add_constr}

The crossing coefficients are constrained by the requirement that the present-day value of the Hubble parameter derived from the deformed luminosity distance remains unchanged. From \cref{eq:DL_def}, we can derive, by differentiating both sides with respect to $z$, the following relation at the present epoch:

\begin{equation}
\left.\frac{d D_L(z)}{dz}\right|_{z=0} = \frac{c}{H_0}\,.
\label{eq:DL_derivative}
\end{equation}

For clarity of notation, we denote the fiducial $\Lambda$CDM luminosity distance $D^{\Lambda CDM}_{L}(z)$ simply as $D_{L}$ throughout this subsection, unless otherwise stated.

Following deformation, the crossed luminosity distance is given by
\[
D_L^{\times}(z) = D_L(z) \times\sum_{i=0}^{N} C_i T_i(\tilde{z}) 
= c (1+z) \int_0^z \frac{dz'}{H^{\times}(z')} \,.
\]
Evaluating this at $z=0$ yields

\begin{equation}
\left.\frac{d D^{\times}_L(z)}{dz}\right|_{z=0} = \left.\frac{d D_L(z)}{dz}\right|_{z=0} \times \sum_{i=0}^{N} C_i T_i(-1) = \frac{c}{H^{\times}_{0}}\,.
\label{eq:DL_cross_derivative}
\end{equation}

Assuming that the deformed model preserves the same present-day Hubble parameter, we should impose the normalization condition $H^{\times}_{0}=H_{0}$. Along with the properties of Chebyshev polynomials and ~\cref{eq:DL_derivative} this leads to the constraint equation
\be
\sum_{i=0}^{N} (-1)^i C_i = 1\,.
\label{eq:consistency}
\ee

Given this constraint, the first-order coefficient $C_1$ is no longer independent 
and can be expressed in terms of the higher-order coefficients as
\[
C_1 = \sum_{i=2}^{N} (-1)^i C_i\,,
\]
which follows directly from the previously derived consistency condition Eq.~\eqref{eq:consistency} and fixing the coefficient $C_0$ is to unity ($C_0 = 1$). The latter is necessary because $C_0$ is completely degenerate with the absolute magnitude $M_B$ in SN Ia observations 
and the sound horizon scale $r_d$ in BAO measurements. 
In other words, any change in $C_0$ can be fully absorbed by adjusting $M_B$ or $r_d$, 
and therefore it does not represent an independent degree of freedom.

Consequently, the additional free parameters in the crossing expansion start from $C_2$, 
and for a polynomial of order $N$, the number of independent degrees of freedom, additional to the fiducial model, is $N-1$.

\subsubsection{Consistency diagnostics and model complexity}
\label{sec:consistency_diagnostics}

In the Crossing Statistics framework, consistency between two probes is assessed by whether they prefer compatible posterior regions in the extended $(\Omega_{m,0}, C_i)$ parameter space, where the background matter density and the crossing coefficients are jointly sampled. Throughout this work, we diagnose consistency primarily through the overlap of the marginalized $1\sigma - 2\sigma$ level in the relevant parameter projections. In particular, such agreement in the $C_i$ space indicates that the datasets favour compatible deformations of the fiducial distance-redshift relation.

In practice, increasing the polynomial order, $N$, introduces additional flexibility that can improve the fit but will eventually lead to overfitting, whereby higher-order parameters are determined primarily by random statistical fluctuations specific to each dataset. Such behaviour can artificially lower the resulting $\chi^2$ and make the contours difficult to interpret in the context of mutual consistency, since they are influenced by each dataset's features, rather than the underlying cosmological model.
To maintain a well-defined notion of consistency, we therefore evaluate the change in $\chi^{2}$, for each crossing order, relative to the $\Lambda$CDM baseline through
\begin{equation}
\Delta \chi^2 \equiv \chi^2_{\Lambda{\rm CDM}} - \chi^2_{T_N}\,,
\label{eq:delta_chi2}
\end{equation}
where $\chi^2_{T_N}$ denotes the minimum $\chi^2$ obtained in the $T_N$ crossing model, i.e.\ the reconstruction including Chebyshev polynomials up to $n_{\max}=N$. 
With the normalization constraint and the choice $C_0=1$, the number of additional degrees of freedom is
\begin{equation}
k = N-1\,,
\label{eq:k_dof}
\end{equation}
corresponding to the free coefficients $\{C_2,\ldots,C_N\}$.

In the limit of large sample size, and for nested models, the statistic $\Delta\chi^2$ is expected to follow a chi-squared distribution with $k$ degrees of freedom, according to Wilks' theorem~\cite{Press2007Numerical}. Thus, when interpreting a particular value of $\Delta\chi^2$ in the context of whether additional flexibility is warranted, we examine the corresponding value of the cumulative distribution function of a chi-squared distribution with $k$ degrees of freedom. This is defined as the probability of obtaining a value for $\Delta\chi^2$ less than or equal to $x$ and is given by~\cite{Press2007Numerical}

\begin{equation}
P(\Delta\chi^2\leq x) \equiv {\rm CDF}_{\rm chi-square}\!\left(x; k\right) = P\!\left(\frac{k}{2},\frac{x}{2}\right)\,,
\label{eq:chi2_cdf}
\end{equation}

where $P(a,x) = \frac{\int_0^xt^{a-1}e^t{\rm d}t}{\Gamma(a)}$ is the regularized lower incomplete gamma function, normalised by the factor $\Gamma(a) = {\int_0^\infty t^{a-1}e^t{\rm d}t}$. In \cref{s:res}, we use the joint behaviour of the posterior overlap in $\{C_i\}$ and the evolution of $\Delta\chi^2$ (i.e. the CDF) with $N$ to identify the range of crossing orders that provides meaningful flexibility without becoming dominated by noise-driven deformations.

Operationally, for each dataset we compute $(\Delta \chi^{2},{\rm CDF})$ at successive crossing orders and track how the probability $P = P(\Delta\chi^2\leq x)$ changes as flexibility is increased. 
Since the crossing extensions are nested parametrizations, as we increase the number of additional degrees of freedom, $\Delta\chi^2$ must increase in absolute magnitude. At the same time, $P$ corresponds to the probability of getting, at random, a $\Delta\chi^2$ at most as large as a particular value, $x$. 
If $\Delta\chi^2$ does not increase sufficiently, given the increase in degrees of freedom, then this probability should decrease, because the value is more compatible with the expected distribution.

A meaningful extension is characterized by a sustained shift toward improvements greater than what is expected from a random distribution (assumed to be a chi-square distribution in this case). Whereas, when further increases in $N$ yield no additional improvement (i.e., saturation or a reversal), then there is superfluous flexibility being introduced by the availability of extra degrees of freedom. We thus define a maximum justified crossing order $N_{max}$ as the point beyond which additional modes no longer produce significant improvements in $\Delta\chi^2$, thus causing the $\{C_{n}\}$ posteriors to overlap simply as a result of poorer constraints on a large number of unnecessary additional parameters.

We emphasize that these statistics are used to determine the maximum justified level of flexibility in the crossing expansion, so that we can assess the mutual consistency of the datasets, rather than to claim evidence for any specific physical features beyond $\Lambda$CDM.
In addition, if the additional degrees of freedom cause non-physical behaviour of the expansion history, we also exclude them (see \cref{s:consistency}).

\subsection{Sampling}

For parameter estimation we employ \texttt{cobaya}~\cite{Torrado:2020xyz}, using its implementation of the Metropolis–Hastings MCMC sampler to explore the posterior distributions of the cosmological and crossing parameters. The SN Ia and BAO likelihoods defined below are treated as statistically independent, and joint constraints are obtained by simply adding their respective log-likelihoods.

The parameter sets sampled for each dataset are:

\[
\Theta_{SN} = \{ \Omega_{m,0},\, M_{ B},\, C_2,\, \ldots,\, C_N \},~~~~~~ \Theta_{BAO} = \{ \Omega_{m,0},\, r_{d},\, C_2,\, \ldots,\, C_N \}.
\]

The priors adopted for all parameters are summarized in Table~\ref{tab:prior}.\\

At each MCMC step, the crossing-modified luminosity distance 
\( D_{L}^{\times}(z) \) 
is evaluated, and the corresponding SN and BAO observables 
\( (m_{B},\, D_{M}/r_{d},\, D_{H}/r_{d},\, D_{V}/r_{d}) \)
are reconstructed, before computing the likelihood. As discussed in~\cref{s:add_constr}, the constraint on the Hubble parameter normalization fixes 
\( C_{1} \) 
in terms of the higher-order coefficients, and $C_0$ is fixed to 1. Thus, only 
\( \{ C_{2}, \ldots, C_{N} \} \) 
are varied independently. Posterior distributions and $1\sigma$ and $2\sigma$ contours shown in the Results section are obtained from the converged MCMC samples after marginalization over the nuisance parameters.\\

We define Gaussian likelihoods using $\Theta_{SN}, \Theta_{BAO}$ as:
\begin{align}
-2 \ln \mathcal{L} = \chi^2 
&=
\left\{
\begin{aligned}
\chi^2_{\mathrm{SN}} &= 
\Delta m_B^{\top} ~\mathrm{C}^{-1}_{\mathrm{SN}} ~\Delta m_B\,, \\[4pt]
\chi^2_{\mathrm{BAO}} &= 
\Delta \mathrm{D}_{\rm BAO}^{\top}\,
\mathrm{C}_{\rm BAO}^{-1}\,
\Delta \mathrm{D}_{\rm BAO}\,,
\end{aligned}
\right.
\end{align}

where $\mathrm{C_X}$ is the full covariance matrix supplied with each dataset X, and 

\begin{align}
\Delta m_{B} &= m_{B}^{\mathrm{data}} - m_{B}^{\times}(\Omega_{m,0}, M_B, C_i)\,,\nonumber\\
\nonumber\\
\Delta \mathrm{D}_{\rm BAO}
&=
\begin{pmatrix}
 D_M / r_d \\
 D_H / r_d \\
 D_V / r_d
 \end{pmatrix}^{\rm data}
 - \,\,\begin{pmatrix}
 D_M / r_d \\
  D_H / r_d \\
 D_V / r_d 
\end{pmatrix}^{\times}
\text{\hspace{-10pt}}(\Omega_{m,0}, r_d, C_i)\,.\nonumber
\end{align}

 The quantities $m_B^{\times}$ for SN Ia and $(D_M/r_d)^{\times},(D_H/r_d)^{\times}, (D_V/r_d)^{\times}$ for BAO  are obtained by evaluating the crossed luminosity distance $D_L^{\times}(z)$  at each MCMC step and converting the result into the corresponding observables in the same format as the original data.\\

\begin{table}[h]
\centering
\small
\begin{tabular}{lccccp{5cm}}
\hline
\hline
\textbf{Dataset} & \textbf{Parameter} & \textbf{Prior}  \\
\hline
 \multirow{2}{*}{Common} & $\Omega_{m,0}$ & $\mathcal{U}$[0.01, 0.99]  \\
~ & $C_{n}$ & $\mathcal{U}$[-5, 5] \\
\hline
SN Ia & $M_{B}$ & $\mathcal{U}$[-30, -15] \\
\hline
BAO & $r_d$ [Mpc] & $\mathcal{U}$[10, 1000] \\
\hline
\hline
\end{tabular}
\caption{Priors}
\label{tab:prior}
\end{table}

\newpage
\section{Results}
\label{s:res}
Within the standard $\Lambda$CDM framework, the BAO and SN Ia datasets already yield mutually consistent constraints on $\Omega_{m,0}$, showing no significant indication of tension or degeneracy. This baseline comparison establishes that both probes, when interpreted through an unmodified $\Lambda$CDM distance-redshift relation, occupy overlapping regions in parameter space at the level of less than $2\sigma$. For completeness, the corresponding best-fit values and $1\sigma$ uncertainties derived from each dataset are summarized in \cref{table:bestfits}.

\begin{table}[h]
\centering
\begin{tabular}{l c}
\hline\hline
\textbf{Dataset} & $\boldsymbol{\Omega}_{m,0}$ \\
\hline\vspace{0.1cm}
DESI DR2 BAO & $0.297^{+0.009}_{-0.008}$ \\
Pantheon+    & $0.332^{+0.019}_{-0.018}$ \\
Union3       & $0.356^{+0.029}_{-0.024}$ \\
\hline
\hline
\end{tabular}
\caption{Best-fit values and $1\sigma$ uncertainties.}
\label{table:bestfits}
\end{table}

However, agreement within $\Lambda$CDM does not guarantee that both datasets prefer the same underlying cosmological trends. There may be discrepancies present which only emerge when additional flexibility is introduced. We therefore employ Crossing Statistics to probe whether the consistency persists once the luminosity distance is allowed to deform away from the fiducial model.

\begin{figure}[htbp]
\centering
\includegraphics[height=.46\textwidth]{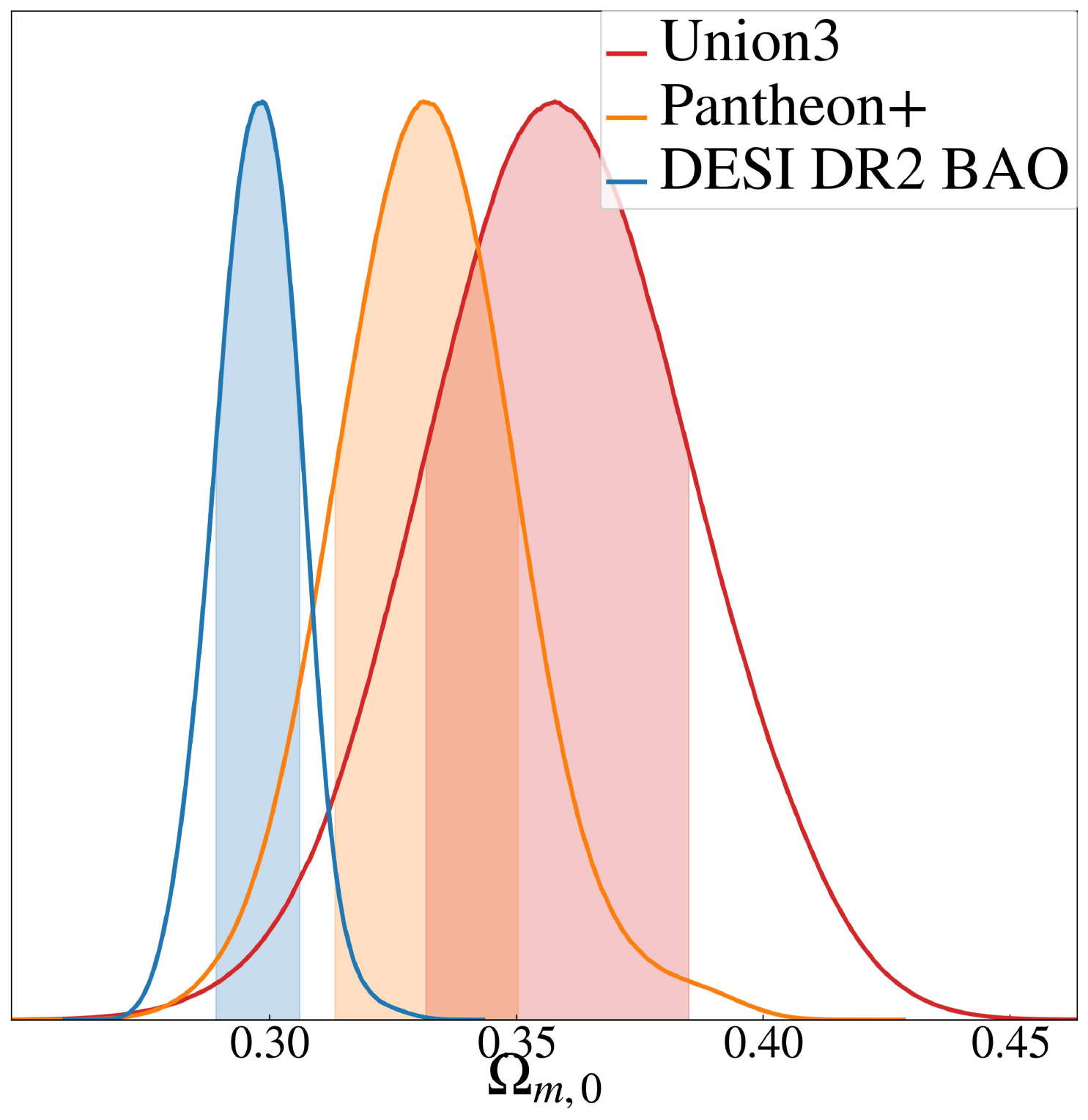}
\qquad
\caption{Marginalized constraints on the present-day matter density parameter $\Omega_{m,0}$ obtained from SN Ia (Pantheon+, Union3) and DESI DR2 BAO within the standard $\Lambda$CDM framework. Shaded regions denote the $1\sigma$ credible intervals. The overlap of the posterior distributions indicates that the two probes are already mutually consistent when interpreted using an unmodified $\Lambda$CDM distance-redshift relation.} \label{fig:Om0_LCDM}
\end{figure}

\subsection{Consistency test}
\label{s:consistency}
If BAO and SN Ia are mutually consistent, they should prefer compatible regions in the extended $(\Omega_{m,0}, C_i)$ parameter space when analysed independently. In particular, consistency manifests as overlapping posterior contours for both the background matter density $\Omega_{m,0}$ and the crossing coefficients $C_i$.

\begin{figure}[htbp]
\centering
\includegraphics[width=.46\textwidth]{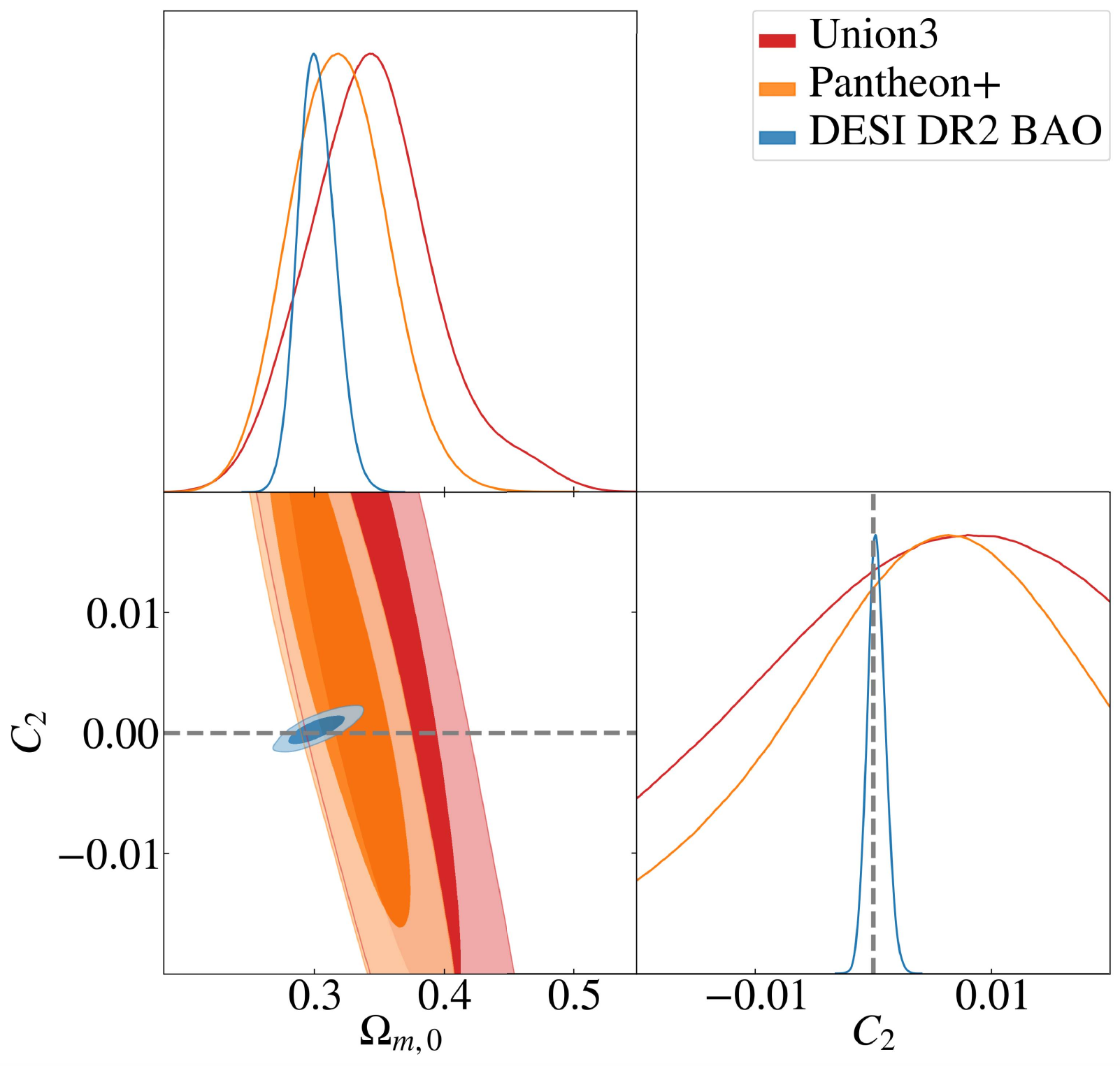}
\qquad
\includegraphics[width=.46\textwidth]{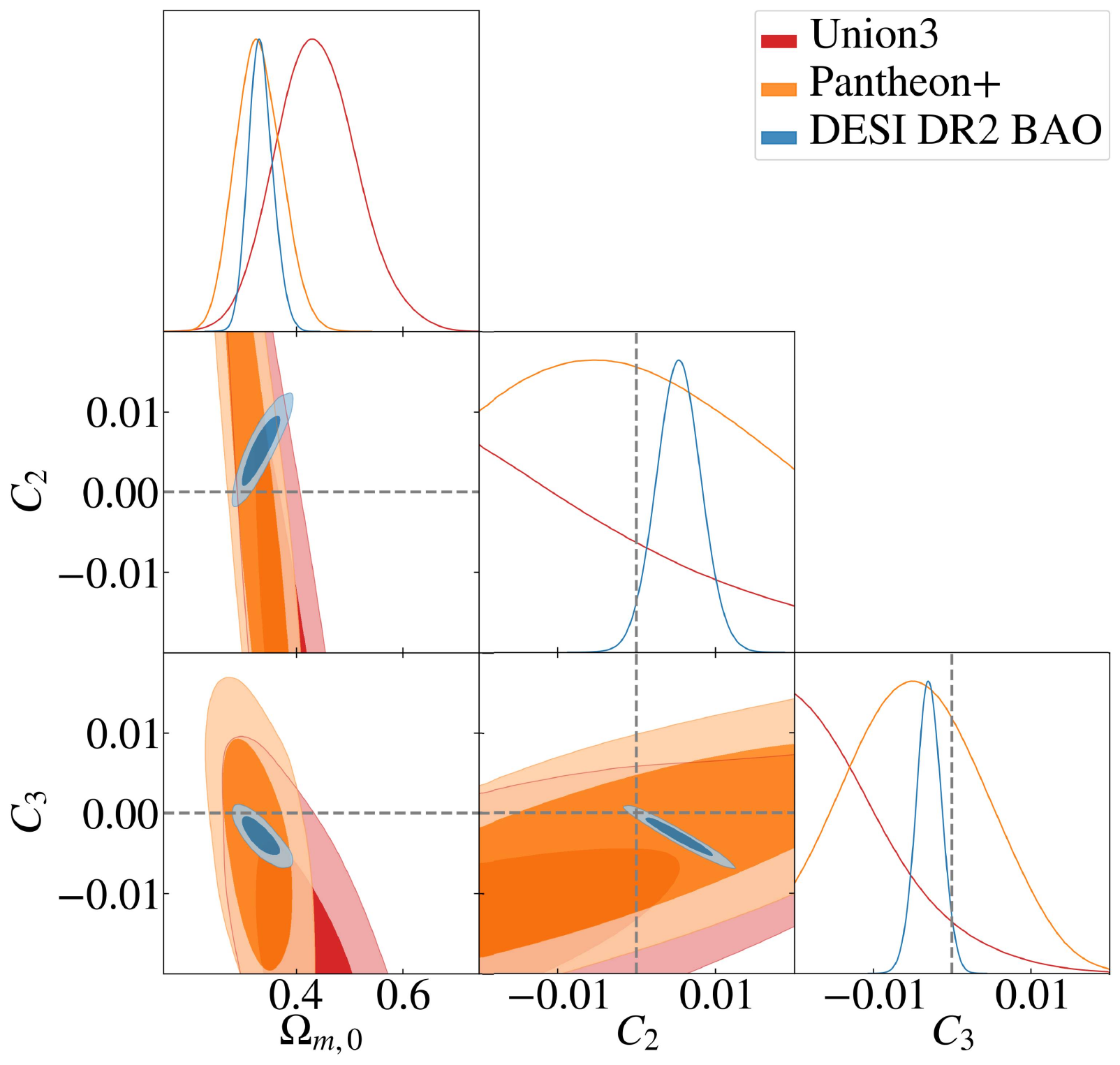}

\caption{Marginalized constraints in the $(\Omega_{m,0}, C_i)$ parameter space obtained from Crossing Statistics analyses of the same datasets. The left panel shows results for the $T_2$ extension (one additional degree of freedom), while the right panel shows the corresponding constraints for the $T_3$ extension (two additional degrees of freedom). Darker and lighter contours indicate the $1\sigma$ and $2\sigma$ level regions, respectively. In both panels, the contours from Union3 SN Ia, Pantheon+ SN Ia, and DESI DR2 BAO exhibit overlap at $1\sigma$, indicating that mutual consistency persists when $\Omega_{m,0}$ is treated as a free parameter and when the crossing order is increased to include $T_2$ and subsequently $T_3$.}
\label{fig:Om0_free_T2_T3}
\end{figure}

\begin{figure}[htbp]
\centering
\includegraphics[height=.6\textwidth]{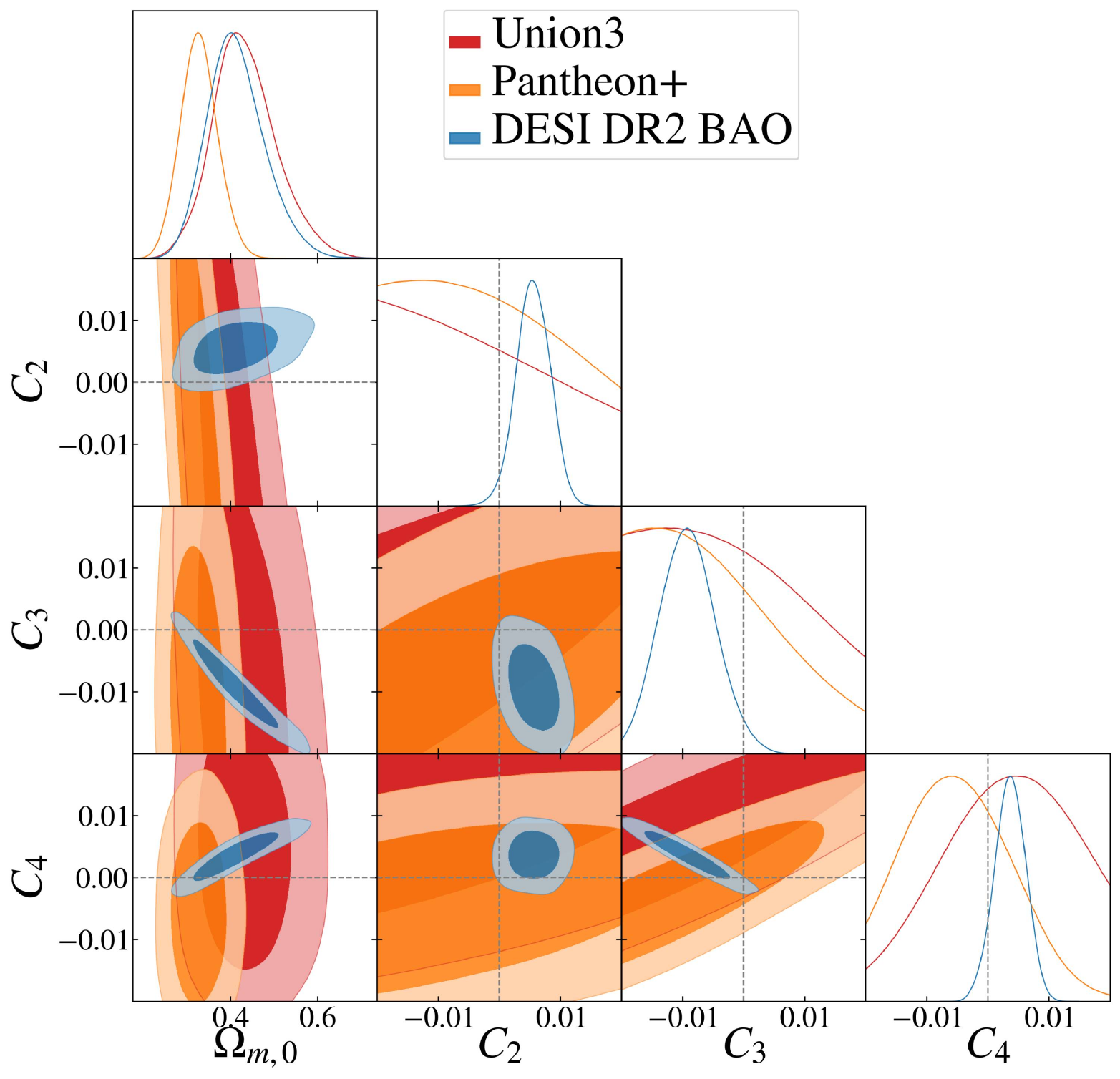}
\qquad
\caption{Marginalized constraints in the extended $(\Omega_{m,0}, C_i)$ parameter space obtained with higher-order crossing extensions. Constraints from DESI DR2 BAO, Pantheon+, and Union3 SN Ia are shown. Black dashed lines indicate the $\Lambda$CDM limit, corresponding to $C_2 = C_3 = C_4 = 0$. Although the contours broaden as additional Chebyshev modes are included, their continued overlap indicates that the datasets remain mutually consistent, with the widening driven primarily by noise-dominated higher-order modes. We also note that the overlapping 1$\sigma$ regions do not contain the $\Lambda$CDM point.}
\label{fig:free_Om0_T4}
\end{figure}

The left and right panels of~\cref{fig:Om0_free_T2_T3} present the constraints in the $(\Omega_{m,0}, C_i)$ plane obtained using Crossing Statistics with one or two additional degree of freedom, respectively. In both cases, the posterior contours of the BAO overlap at $1\sigma$ - $2\sigma$ with each of the SN Ia datasets, indicating that the consistency observed for fixed $\Omega_{m,0}$ is preserved after marginalization over different values and as additional Chebyshev modes are included. As shown in~\cref{fig:free_Om0_T4}, the SN Ia and BAO datasets continue to exhibit regions of overlap in the $1\sigma$ contours in all $(\Omega_{m,0}, C_i)$ projections. This means each SN Ia dataset is consistent with the BAO, within the Crossing Statistics framework.

A recurring feature visible in~\cref{fig:Om0_free_T2_T3,fig:free_Om0_T4} are the broader posterior contours obtained from SN Ia relative to BAO. This difference arises from the distinct propagation of observational uncertainties into the luminosity-distance space: magnitude errors enter logarithmically for SN Ia, while BAO observables propagate linearly. Consequently, SN Ia data allow a wider range of crossing-induced deformations, naturally producing broader $C_i$ constraints.

We restrict the analysis to crossing orders $N \leq 5$ in general, since reconstructions at higher order begin to exhibit non-physical behaviour in the luminosity-distance relation when only information from individual datasets are used. 
When additional crossing coefficients are introduced beyond this order, the deformations tend to overfit the data, increasingly tracking statistical fluctuations from within each dataset. 
In these cases the reconstructed expansion history leads to a non-monotonic luminosity distance $D_L(z)$ evolution in the higher redshift range of the data, which we consider to be unphysical. In the case of Pantheon+ this occurs even at $T_5$, or 4 additional degrees of freedom. This behaviour signals that the degree of allowed flexibility is too high, and so we consider crossing orders $N>5$ ($N>4$ for Pantheon+) to be dominated by noise-driven overfitting and exclude them from the analysis entirely, on physical grounds.

For the SN Ia datasets, the posterior contours remain sufficiently broad that the $\Lambda$CDM point ($C_i = 0$ for all $i$) is consistently contained within the $1\sigma$ region. In contrast, the BAO (DESI DR2) constraints exhibit a displacement of the region of strongest preference away from the undeformed $\Lambda$CDM limit. The peak of the BAO posterior is therefore not centered exactly at $C_i = 0$, although the $\Lambda$CDM point remains just within the broader $2\sigma$ region.
This indicates that while both probes are statistically consistent within the crossing framework, the regions of best agreement between both pairs of datasets show a slight preference for non-zero deformation parameters, reaching a level of tension of $1\sigma-2\sigma$ with the $\Lambda$CDM solution.

\begin{figure}[htbp]
\centering
\includegraphics[height=.35\textwidth]{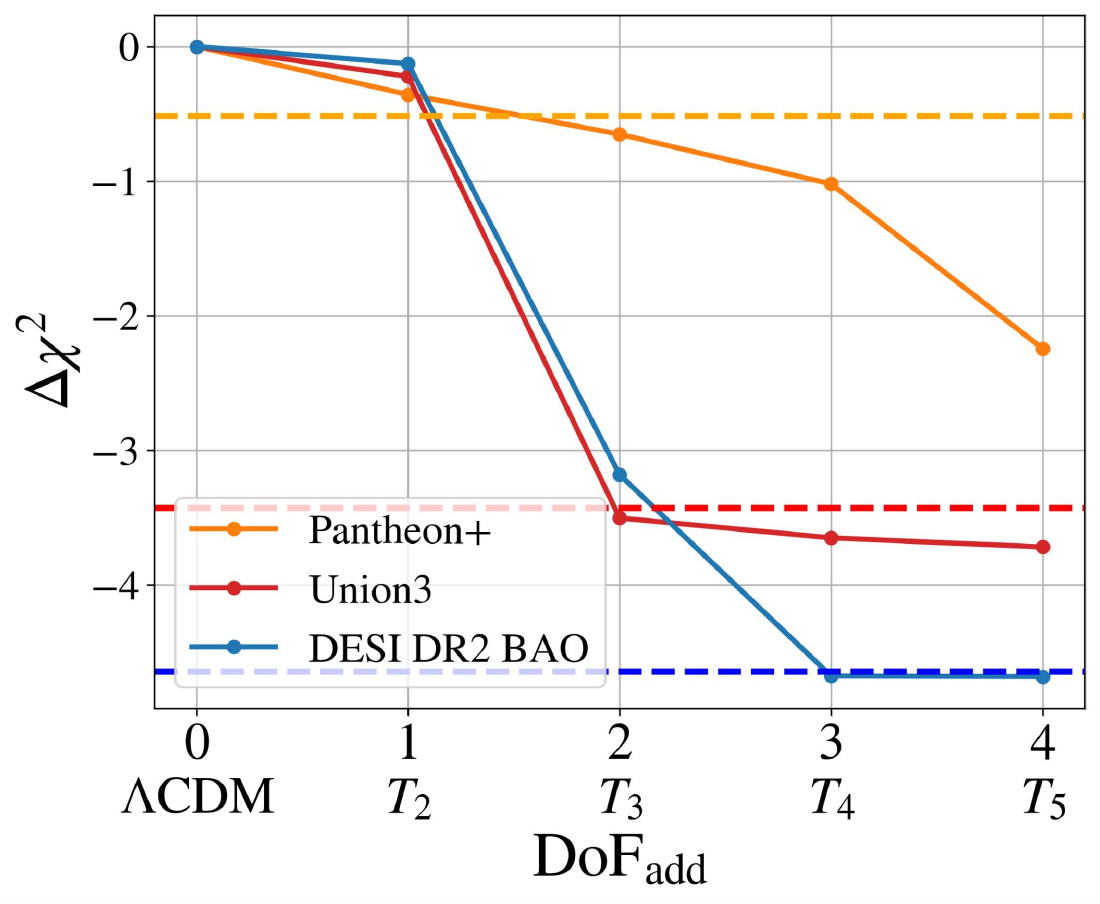}
\hspace{0\textwidth}
\includegraphics[height=.35\textwidth]{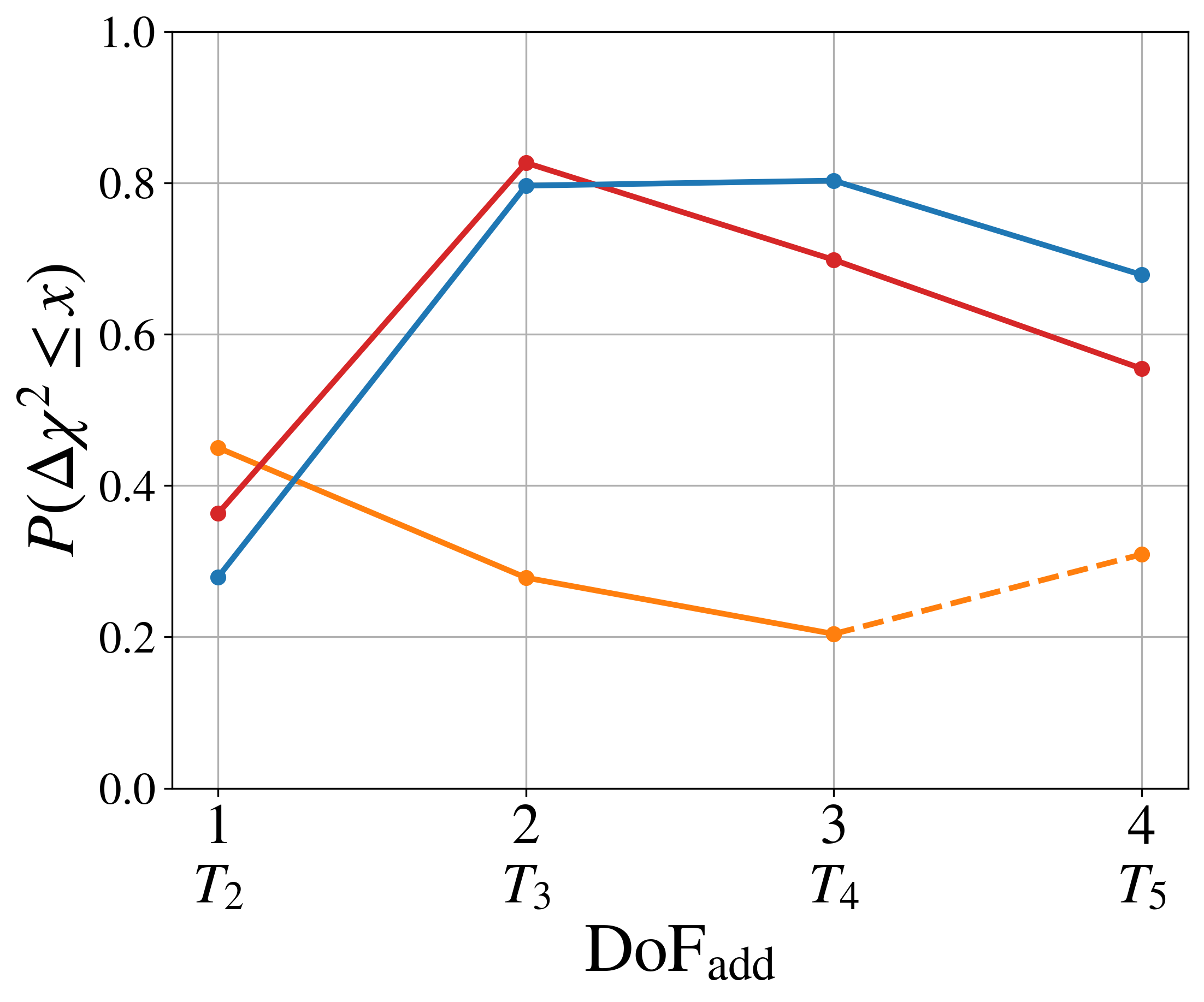}
\caption{\textit{Left:} The solid curves show $\Delta\chi^2$ for each dataset (Pantheon+, Union3, and DESI DR2) relative to the $\Lambda$CDM baseline as the crossing order increases. 
The horizontal dashed lines indicate the corresponding $\Delta\chi^2$ values obtained from the best fit $w_0w_a$CDM model for each dataset. 
\textit{Right:} Cumulative distribution function $P(\Delta\chi^2 \leq x)$ for the improvement in $\chi^2$ obtained from crossing reconstructions as a function of the number of additional degrees of freedom (and crossing order).
The dashed orange line for Pantheon+ indicates that in the case of $T_5$, the seemingly improved fit actually arises from a non-physical solution that overfits to the data.
}
\label{fig:free_Om0_chi2}
\end{figure}

As part of the consistency test, we also examine the $\chi^2$ values as a function of the number of crossing degrees of freedom.
In \cref{fig:free_Om0_chi2}, for all datasets except Pantheon+, the largest improvement occurs when two additional crossing parameters are introduced. Beyond this point the statistical gain rapidly diminishes, as additional crossing parameters give an unnecessarily high level of freedom to the expansion history. This means that consistency may be trivially achieved simply by the resultant broadening of the contours.

In summary, even when additional crossing freedom is allowed, the DESI DR2 BAO constraints remain consistent with those obtained from each of the supernova datasets individually at the $2\sigma$ level. We thus conclude that the preferred distance-redshift behaviour is mutually compatible across probes, independent of the background cosmological model.

\subsection{Combined reconstructions}

Having established that the SN Ia and BAO datasets are mutually consistent within the Crossing Statistics framework, it is justified to consider their joint use for reconstruction purposes. Therefore, we subsequently combine DESI DR2 BAO with each supernova sample. 

Since the datasets show a preference for consistent deviations from
the fiducial $\Lambda$CDM distance-redshift relation, the combined analysis allows us to explore how the mutually preferred deformations translate into modifications of the expansion history and related cosmological quantities.

As discussed previously, in order to determine an appropriate truncation order, we examine the improvement in the $\chi^2$ as additional crossing parameters are introduced. Specifically, we evaluate $\Delta\chi^2$ relative to the fiducial model and compute the CDF of the corresponding expected chi-square distribution, which depends on the additional degrees of freedom. We find that beyond the $T_3$ model, the reduction in $\chi^2$ becomes statistically insignificant. In particular, once two additional degrees of freedom are included, further extensions yield only marginal improvement in $\chi^2$, so the corresponding $\chi^2$ CDF indicates a decreasing validity of higher-order models. This behaviour suggests that two additional degrees of freedom are sufficient to capture the deformation preferred by the data, and higher-order terms primarily introduce unnecessary flexibility. Moreover, at higher polynomial order, the reconstructed $D_L(z)$ once again begins to exhibit unphysical behaviour. 
We therefore restrict the reconstruction analysis to the crossing models of order $T_3$ and lower, which allows for flexibility in capturing the deformations preferred by the combined data, while avoiding unnecessary additional degrees of freedom.

\begin{figure}[htbp]
\centering
\includegraphics[height=.265\textwidth]{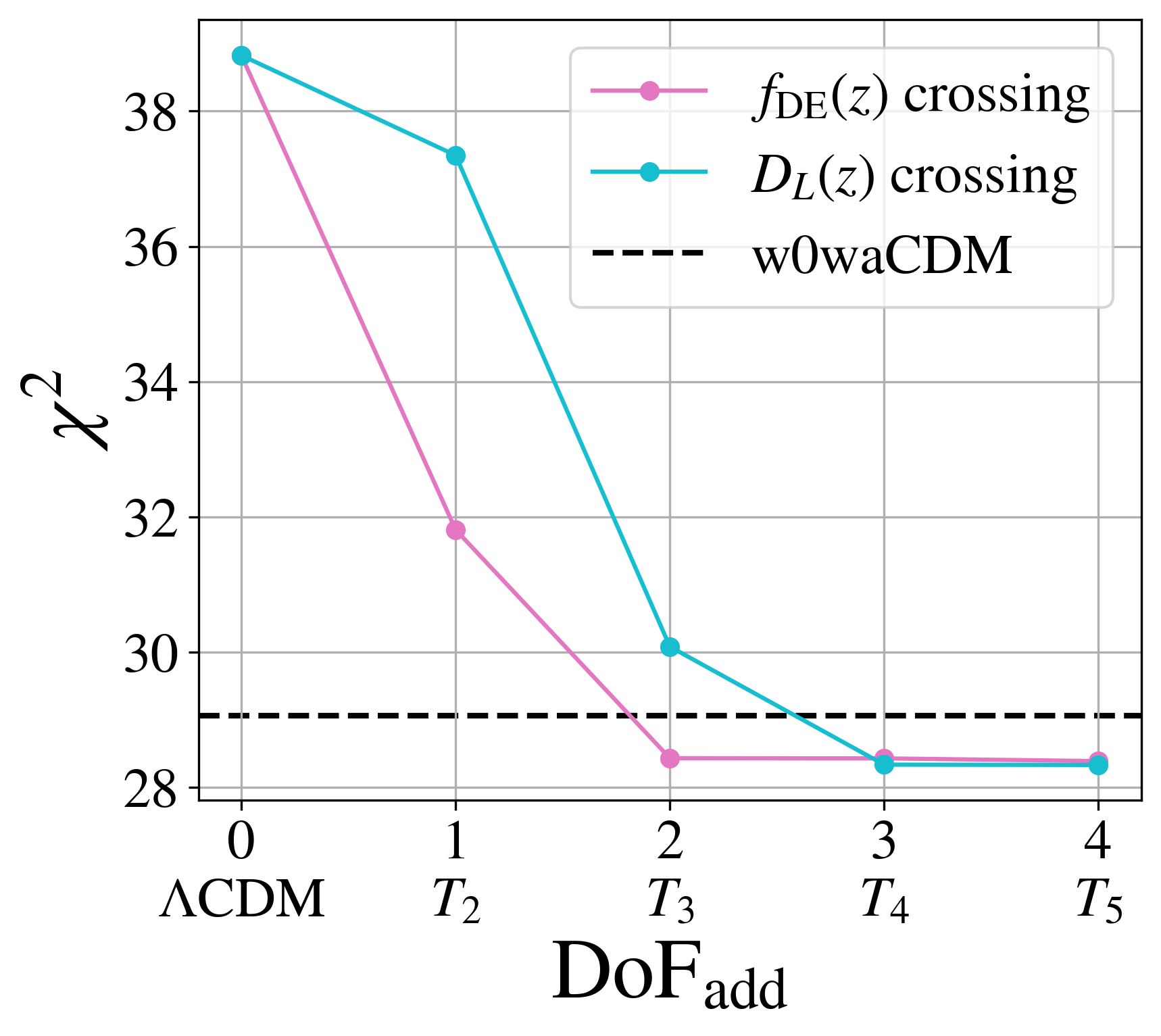}
\includegraphics[height=.265\textwidth]{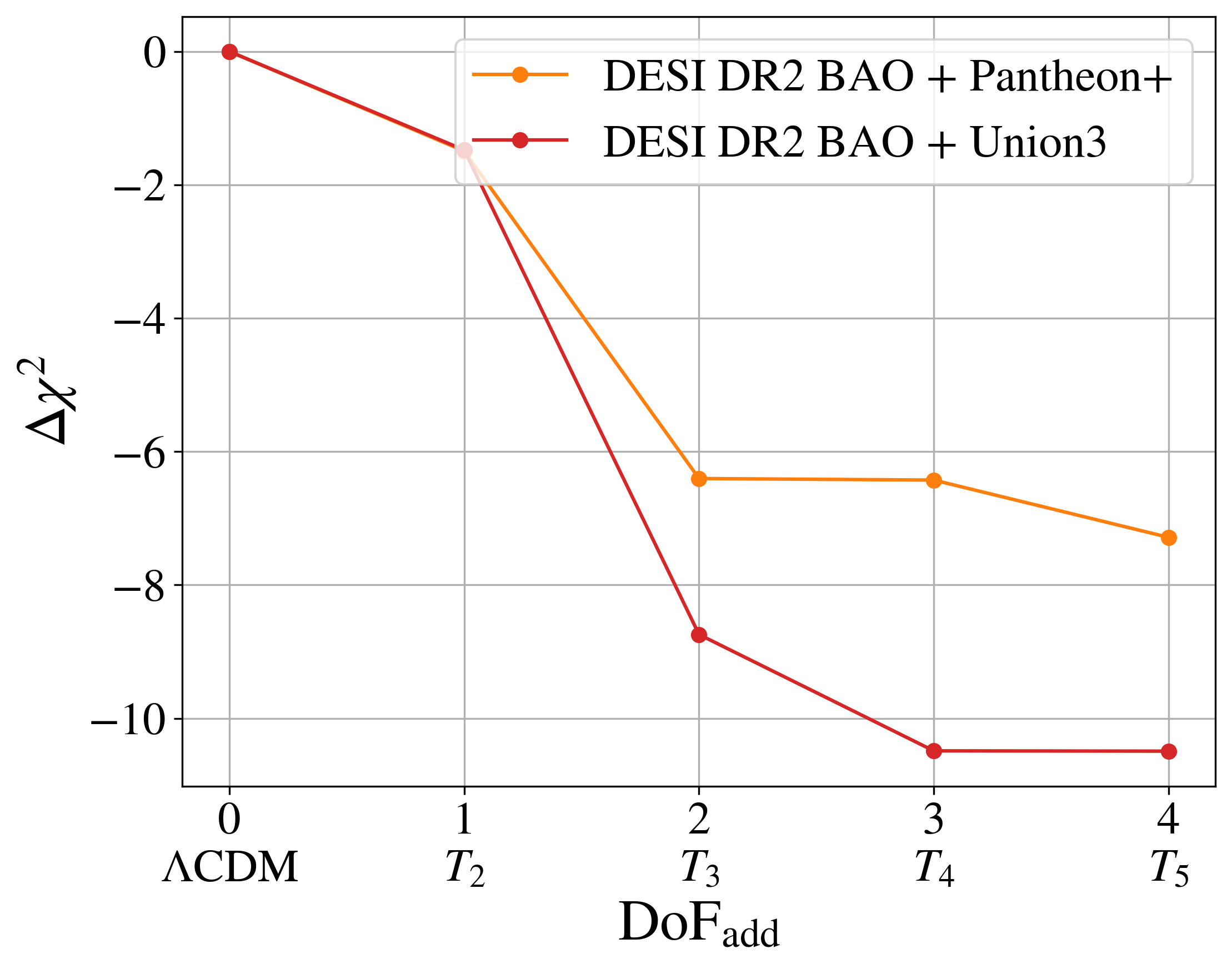}
\includegraphics[height=.265\textwidth]{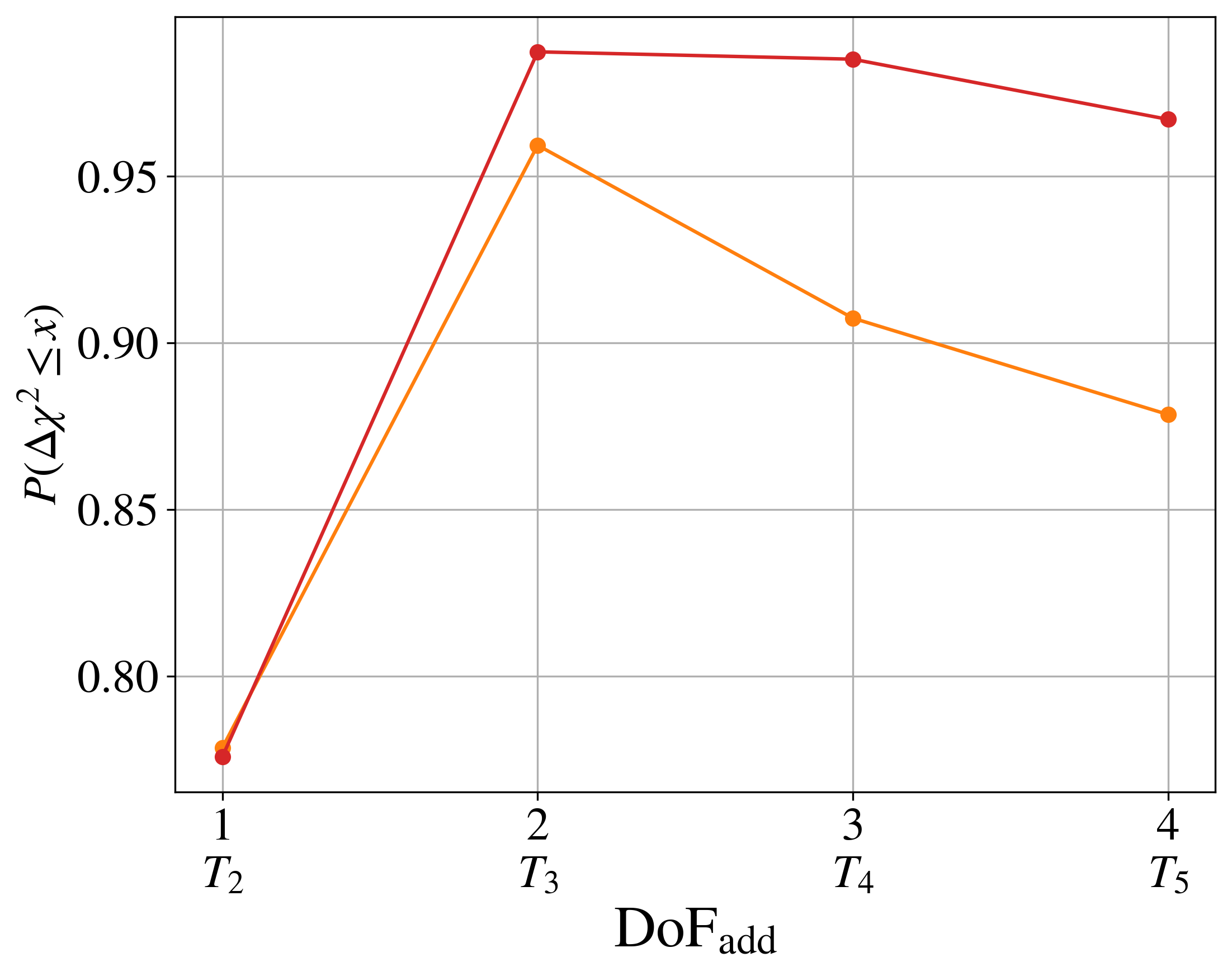}
\caption{
\textit{Left:} Best fit $\chi^2$ values, from the combination of Union3 and DESI DR2 BAO, as a function of the number of additional degrees of freedom (${\rm DoF}_{\rm add}$), corresponding to increasing crossing order from $\Lambda$CDM ($T_0$) to $T_5$. 
Cyan points correspond to crossing applied to the luminosity distance $D_L(z)$, while pink points denote the results obtained by applying crossing to $f_{\rm DE}(z)$. 
The dashed black line indicates the $\chi^2$ value of the best fit $w_0w_a$CDM model for comparison. 
\textit{Middle:} Improvement in $\chi^2$, $\Delta\chi^2 = \chi^2_{\Lambda{\rm CDM}} - \chi^2_{T_N}$, for crossing applied to $D_L(z)$ and the combined data, as a function of the number of $\mathrm{DoF}_{\mathrm{add}}$. \textit{Right:} Corresponding CDF values, assuming a chi-squared distribution with $k = N-1$ $\mathrm{DoF}_{\mathrm{add}}$. 
The saturation of the change in $\chi^2$ beyond two crossing parameters indicates that higher-order extensions are not providing any meaningful additional freedom.}
\end{figure}

\begin{figure}[htbp]
\centering
\includegraphics[height=.8\textwidth]{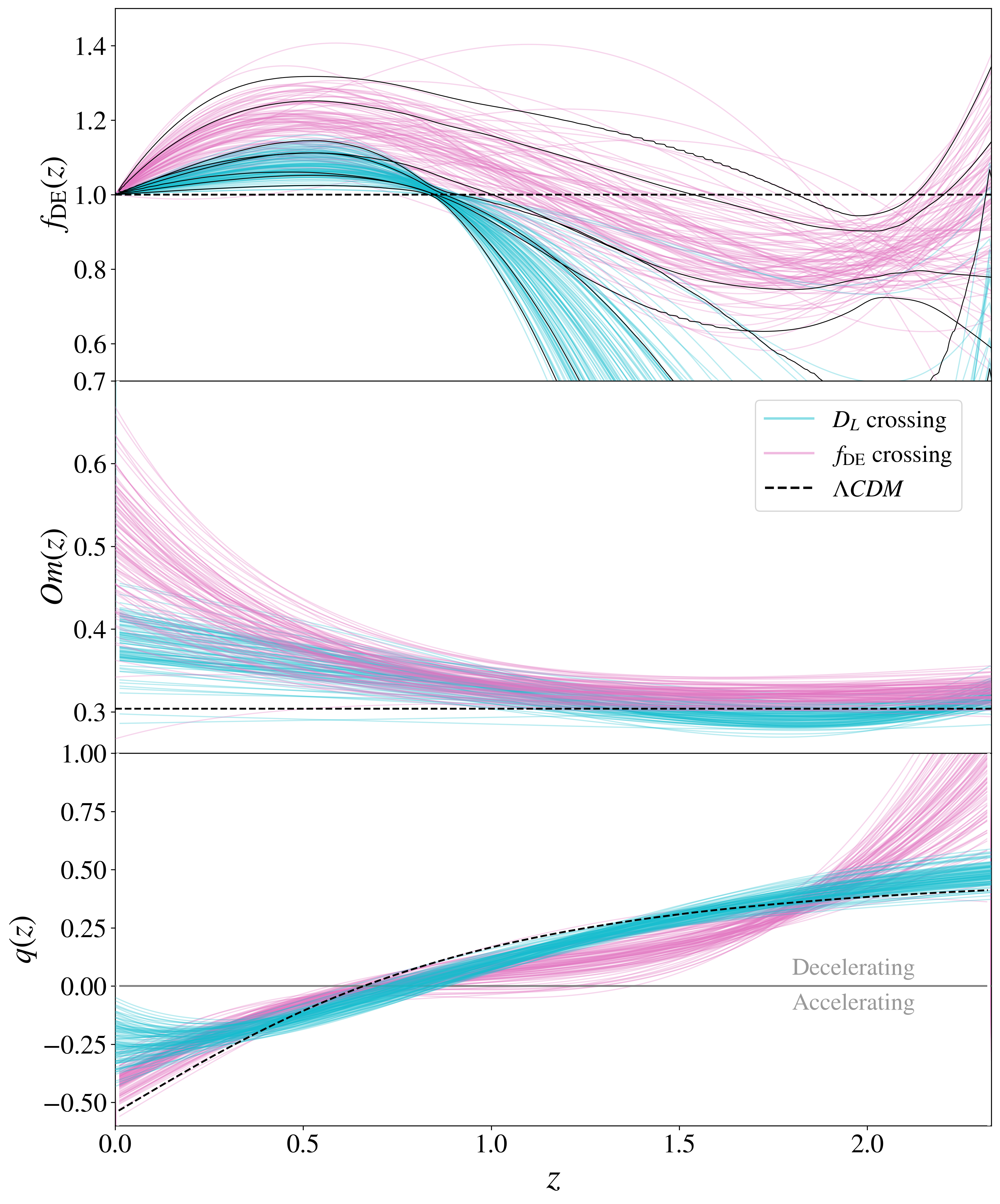}
\caption{Reconstructed cosmological functions obtained within the Crossing Statistics framework using the $T_3$ extension (corresponding to two additional independent crossing degrees of freedom, $C_2$ and $C_3$). From top to bottom we show: the reconstructed relative dark energy density function $f_{\rm DE}(z)$, the $Om(z)$ diagnostic, and the deceleration parameter, $q(z)$. 
In each panel, the dashed black line corresponds to the $\Lambda$CDM best fit, 
while the solid black lines in the top panel indicate the regions encompassed within $1\sigma$ and $2\sigma$ confidence intervals of the sampled crossing parameters. 
Blue curves correspond to crossing applied to the luminosity distance $D_L$, whereas orange curves represent realizations obtained by applying crossing to $f_{\rm DE}$.
The spread of the reconstructed trajectories illustrates the allowed deformation patterns under the adopted crossing order.}
\label{f:recon}
\end{figure}

We use the posterior samples of the crossing parameters in the $D_L$ space, fit to the combination of DESI DR2 BAO and Union3, to reconstruct the corresponding evolution of $D_L$ with redshift. From this we can infer the corresponding behaviour of the relative dark energy density, $f_{\rm DE}(z)$. For each MCMC sample, the crossing-modified expansion history is converted into $H(z)$ and subsequently into $f_{\rm DE}(z)$ using the background Friedmann relation~\cref{eq:friedmann_fde}.
The reconstructed $f_{\rm DE}(z)$, corresponding to the cyan curves in \cref{f:recon}, exhibits deviations from the constant behaviour expected in $\Lambda$CDM, indicating an evolving dark energy density within the allowed parameter space. At high redshift, the preferred behaviour is highly varied and the reconstructions display larger excursions from the constant $\Lambda$CDM value. This behaviour can be attributed to the relative sparsity of data points in this regime, which affords higher-order crossing modes more flexibility. In addition, the absence of CMB information in the present analysis implies that early-Universe constraints on the expansion history are not imposed, further increasing the allowed freedom at high redshift.

To test the robustness of the reconstruction, we perform an additional crossing analysis directly in the $f_{\rm DE}$ space, following an approach analogous to that adopted in the DESI analysis~\cite{DESI:2025fii}. In \cref{f:recon}, these reconstructions are represented by pink curves.
We then compare the resulting $f_{\rm DE}(z)$ reconstruction with that obtained from crossing applied in the observable $D_L(z)$ space.  Since the crossing deformation is implemented in different functional spaces, one should not expect the two reconstructions to coincide exactly. The flexibility introduced by a finite set of Chebyshev modes propagates differently in $D_L(z)$ and in $f_{\rm DE}(z)$, leading  to quantitative differences in general, particularly in regions where the data constraints are weak. Nevertheless, if the underlying cosmological information encoded in the data is consistent, the reconstructed dark energy evolution should remain statistically compatible.

While the behaviour at higher redshift differs due to the increased flexibility in poorly constrained regions, the two reconstructions exhibit similar qualitative trends at low redshift. In particular, for $z \lesssim 1$, both approaches indicate an enhancement of the relative dark energy density compared to the $\Lambda$CDM expectation $f_{\rm DE}(z)=1$, with overlapping $1\sigma$ regions and consistency at better than $1\sigma$. At intermediate redshifts ($z \gtrsim 1$), the reconstructions tend to prefer values of $f_{\rm DE}(z)$ below unity, although the uncertainties grow substantially as the data constraints weaken. This is because, at higher redshifts where matter dominates, the relative contribution of dark energy to the expansion history is comparatively negligible. This transition from $f_{\rm DE}(z)>1$ at low redshift to $f_{\rm DE}(z)<1$ at higher redshift is observed in both reconstruction spaces, despite quantitative differences arising from the different crossing implementations. 

The corresponding $Om(z)$ diagnostic departs from the constant behaviour 
expected in $\Lambda$CDM. Instead of remaining fixed at $\Omega_{m,0}$, the reconstructed curves tend to increase with decreasing redshift.

The reconstructed deceleration parameter $q(z)$ shows a transition 
from positive values at higher redshift (decelerating expansion) toward negative values at low redshift (accelerating expansion). While the qualitative behaviour 
remains consistent with the standard expansion picture, the reconstructed 
curves allow a somewhat different class of acceleration histories than those 
predicted by $\Lambda$CDM or the reconstruction from crossing on $f_{\rm DE}(z)$.

\section{Discussion and Conclusions}
\label{s:conc}

Across all analyses, the BAO (DESI DR2) and SN Ia (Pantheon+, Union3) datasets exhibit mutually consistent behaviour in the $(\Omega_{m,0}, C_i)$ parameter space, for $\Lambda$CDM and its Crossing Statistics extensions ($T_N$). The marginalized posterior distributions obtained from the different probes overlap at the $1\sigma - 2\sigma$ level, indicating statistically compatible luminosity-distance trends, even once additional deformation freedom is introduced. 

The consistency test was further examined using the evolution of the best fit $\chi^2$ as a function of crossing order. We find that the largest improvement occurs when up to two additional degrees of freedom (corresponding to the $T_3$ extension) are included. Beyond this, the reduction in $\chi^2$ becomes marginal. Above $T_5$, the reconstructions of the luminosity-distance relation begin to exhibit unphysical features, such as oscillatory behaviour or non-monotonic distance evolution. For this reason, we restrict the analysis to low-order crossing extensions ($N \leq 3$), where the number of additional degrees of freedom are statistically justified and the solutions remain physical.

Having established consistency, we combine DESI DR2 BAO and Union3 data to reconstruct the relative dark energy density $f_{\rm DE}(z)$. Crossing deformations are applied in both the observable $D_L(z)$ space and directly in the $f_{\rm DE}(z)$ space to test robustness. Because crossing is implemented in different functional spaces, exact agreement between the two reconstructions is not expected. The finite Chebyshev expansion propagates differently in $D_L$ and $f_{\rm DE}$, particularly at high redshift where data are sparse.

Nevertheless, at $z \lesssim 1$, the reconstructed $f_{\rm DE}(z)$ functions overlap within the 1$\sigma$ region, and similar trends are observed in the derived kinematic quantities, the $Om(z)$ diagnostic and the deceleration parameter $q(z)$. At higher redshift, the uncertainties increase substantially due to the limited number of data points, the weaker effect of dark energy on the expansion history and the absence of CMB priors, leading to larger deformations in the reconstructed functions. Overall, the reconstruction suggests departures of more than $2\sigma$ from constant-$\Lambda$ behaviour at various regions across redshift.

The Crossing Statistics methodology thus provides a controlled and flexible framework for probing departures from $\Lambda$CDM, directly at the level of cosmological observables, enabling consistency tests and reconstruction analyses without assuming a specific dark energy parametrization.
Applying this framework to current datasets, we find that the SN Ia (Pantheon+, Union3) and BAO (DESI DR2) measurements remain mutually consistent within the extended $(\Omega_{m,0}, C_i)$ parameter space, with their posterior contours overlapping at the 1-2$\sigma$ level. The combined analysis reveals reconstructions of the expansion history that are broadly consistent across different crossing implementations. In particular, reconstructions performed in both $D_L(z)$ and $f_{\rm DE}(z)$ spaces exhibit similar evolutionary trends at low redshift, deviating from the $\Lambda$CDM behaviour at more than $2\sigma$.

\appendix

\section{Consistency test including DES Supernovae}
\label{a:DES}

As discussed in the main text, the DES-Y5 supernova sample, from \cite{DES:2024jxu}, was excluded from the primary analysis due to its significantly lower maximum redshift compared to the other datasets. 
This mismatch in redshift coverage complicates the application of a common normalization in the crossing-statistics framework, leading to difficulties in making a fair comparison.

During the course of this work, the DES-Dovekie supernova compilation~\cite{DES:2025sig} was released, providing an updated version of the DES-Y5 dataset with improved calibration. We include both versions in this analysis, where we compare them to the DESI DR2 BAO, as was done with Pantheon+ and Union3 in the primary analysis. 

To address the mismatch in redshift ranges, we restrict the analysis to a similar redshift range as the DES supernovae,   $z_{\max}^{\rm DES\text{-}Y5} \simeq 1.12$ and $z_{\max}^{\rm DES\text{-}Dovekie} \simeq 1.144$. In practice, we remove the highest-redshift DESI DR2 BAO points, so that the maximum redshift remaining is $z_{\max}^{\rm DESI} \simeq 1.321$, ensuring a more comparable redshift range and hyperparameter definition.

Following the same procedure as in \cref{s:res}, we systematically increase the allowed deformation freedom through the addition of crossing parameters, and examine the preferred regions of the hyperparameter space, taking into account the improvement in $\chi^2$. 

Based on the latter, we may add a maximum of two crossing parameters to the fiducial $\Lambda$CDM model. Beyond this point, the improvement in $\chi^2$ does not justify the additional degrees of freedom, leading to inflated posteriors that are trivially consistent.

\begin{figure}[htbp]
\centering
\includegraphics[height=.4\textwidth]{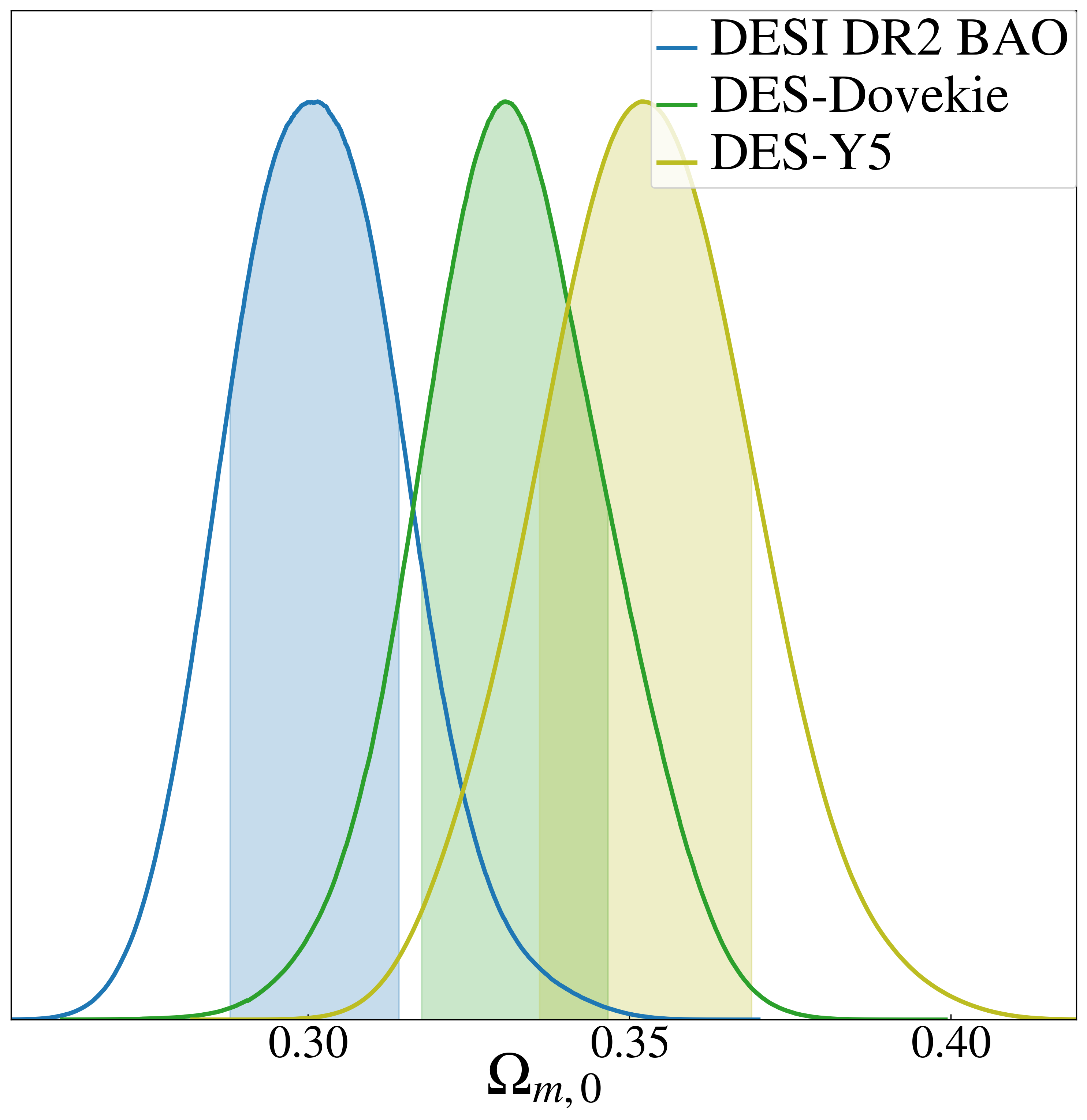}
\hspace{0\textwidth}
\includegraphics[height=.4\textwidth]{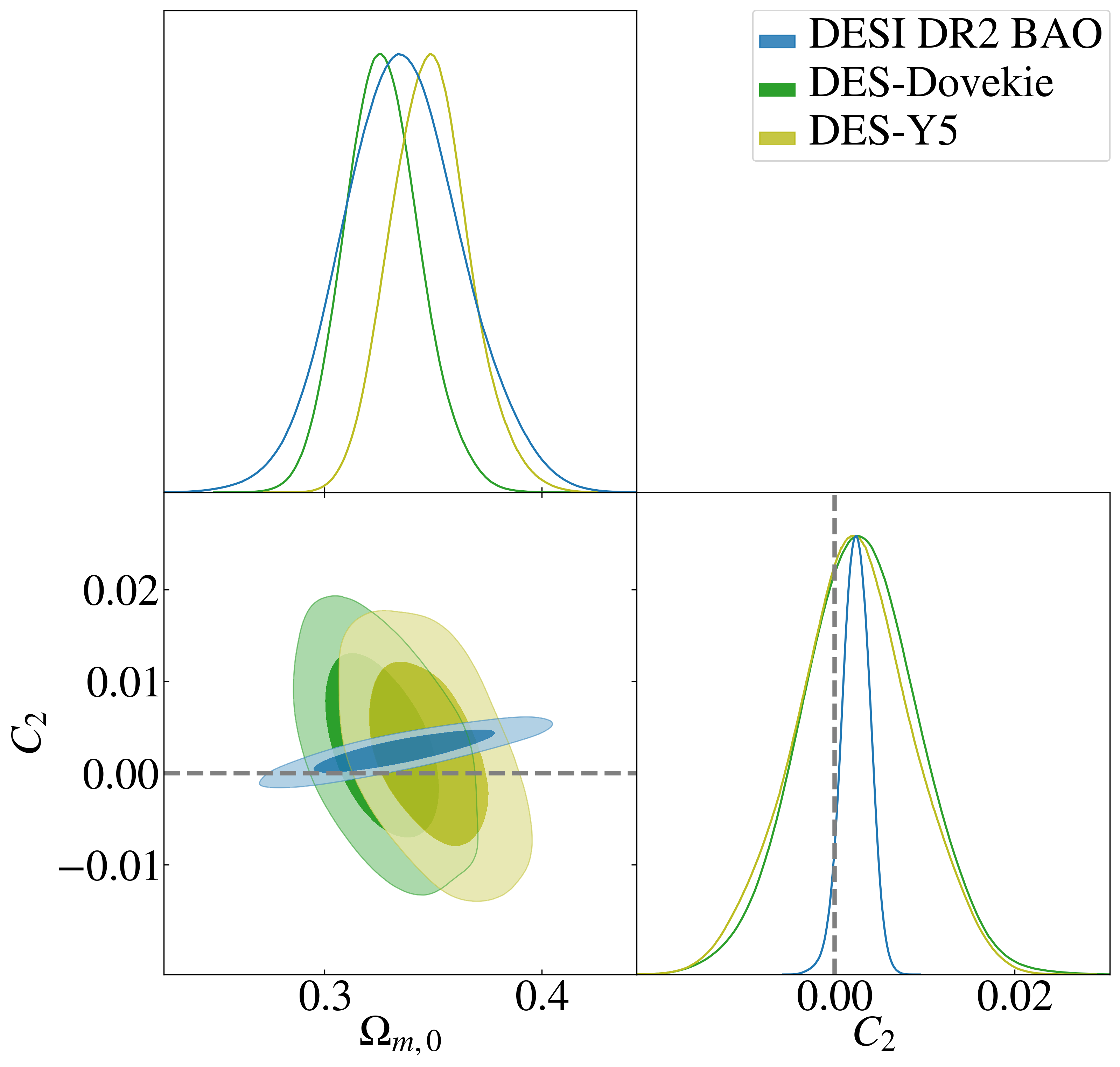}
\caption{\textit{Left:} Marginalized constraints on the present-day matter density parameter $\Omega_{m,0}$, obtained from SN Ia (DES-Dovekie, DES-Y5) and DESI DR2 BAO, within the standard $\Lambda$CDM framework. Shaded regions denote the $1\sigma$ level. The overlap of the posterior distributions indicates that the two probes are already mutually consistent when interpreted using an unmodified $\Lambda$CDM distance-redshift relation, at $2\sigma$ level.
\textit{Right:} Marginalized constraints, in the $(\Omega_{m,0}, C_2)$ parameter space, obtained from the Crossing Statistics analysis of the same datasets. Darker and lighter contours indicate the  $1\sigma$ and $2\sigma$ level regions, respectively. The contours from DES-Dovekie, DES-Y5, and DESI DR2 BAO exhibit overlap at $1\sigma$ in the space of hyperparameters, when the crossing order is increased to $T_2$.
}
\label{DES-crossing}
\end{figure}

In \cref{DES-crossing}, we show separately the posteriors from $\Lambda$CDM and the crossing extension with $(\Omega_{m,0},C_2)$. All three datasets exhibit mutually consistent constraints to within $2\sigma$. In the case of crossing with $C_2$, the region of best agreement is only consistent with $\Lambda$CDM ($C_2 = 0$) at less than about $1\sigma$, for reasonable values of $\Omega_{m,0}$. The overlapping region, with the DESI DR2 BAO, of DES-Dovekie is only marginally more consistent with $\Lambda$CDM than that of DES-Y5. 
At the next order in crossing, where $T_2$ and $T_3$ are included, the posteriors of the DES-Dovekie data are slightly more consistent than DES-Y5, with respect to DESI DR2 BAO, at just over $2\sigma$. However, this more flexible reconstruction exhibits non-physical behaviour, reflecting the onset of overfitting to each dataset's statistical fluctuations. Practically, this means that these disagreements between the datasets in the crossing extension up to $T_3$, while perhaps interesting, do not reflect anything meaningful about the consistency. The maximum possible degrees of freedom, where we can apply the crossing framework to the low redshift data, is for the extension with $(\Omega_{m,0},C_2)$.

We therefore conclude that, for the available physically meaningful crossing extensions, the DES-Y5 and DES-Dovekie datasets are both consistent with the low-redshift data from DESI DR2 BAO, at $1\sigma - 2\sigma$. This is in agreement with the relevant results found for the other supernova datasets. The mutually preferred region of the DESI DR2 BAO and DES-Dovekie seems marginally more consistent with $\Lambda$CDM than that of DESI DR2 BAO and DES-Y5, but both at $\gtrsim1\sigma$. 

\FloatBarrier

\bibliographystyle{JHEP}
\bibliography{refs}

@article{DESI:2025fii,
    author = "Lodha, K. and others",
    collaboration = "DESI",
    title = "{Extended dark energy analysis using DESI DR2 BAO measurements}",
    eprint = "2503.14743",
    archivePrefix = "arXiv",
    primaryClass = "astro-ph.CO",
    reportNumber = "FERMILAB-PUB-25-0164-PPD",
    doi = "10.1103/w4c6-1r5j",
    journal = "Phys. Rev. D",
    volume = "112",
    number = "8",
    pages = "083511",
    year = "2025"
}

@article{DESI:2025zgx,
    author = "Abdul Karim, M. and others",
    collaboration = "DESI",
    title = "{DESI DR2 Results II: Measurements of Baryon Acoustic Oscillations and Cosmological Constraints}",
    eprint = "2503.14738",
    journal={},
    archivePrefix = "arXiv",
    primaryClass = "astro-ph.CO",
    reportNumber = "FERMILAB-PUB-25-0169-PPD",
    month = "3",
    year = "2025"
}

@article{SupernovaSearchTeam:1998fmf,
    author = "Riess, Adam G. and others",
    collaboration = "Supernova Search Team",
    title = "{Observational evidence from supernovae for an accelerating universe and a cosmological constant}",
    eprint = "astro-ph/9805201",
    archivePrefix = "arXiv",
    doi = "10.1086/300499",
    journal = "Astron. J.",
    volume = "116",
    pages = "1009--1038",
    year = "1998"
}

@article{Perlmutter:1999jt,
  author = {Perlmutter, S. and others},
  title = {Measurements of Omega and Lambda from 42 High-Redshift Supernovae},
  journal = {Astrophysical Journal},
  volume = {517},
  pages = {565--586},
  year = {1999},
  eprint = {astro-ph/9812133},
  archivePrefix = {arXiv},
  doi = {10.1086/307221}
}

@article{previous_BAO_measurement,
  author = {Eisenstein, D. J. and others},
  title = {Detection of the Baryon Acoustic Peak in the Large-Scale Correlation Function of SDSS Luminous Red Galaxies},
  journal = {Astrophysical Journal},
  volume = {633},
  pages = {560--574},
  year = {2005},
  eprint = {astro-ph/0501171},
  archivePrefix = {arXiv},
  doi = {10.1086/466512}
}

@ARTICLE{2007GReGr..39.1047E,
       author = {{Ellis}, George F.~R.},
        title = "{On the definition of distance in general relativity: I. M. H. Etherington (Philosophical Magazine ser. 7, vol. 15, 761 (1933))}",
      journal = {General Relativity and Gravitation},
         year = 2007,
        month = jul,
       volume = {39},
       number = {7},
        pages = {1047-1052},
          doi = {10.1007/s10714-006-0355-5},
       adsurl = {https://ui.adsabs.harvard.edu/abs/2007GReGr..39.1047E}
}

@article{Ellis:2007,
  author = {Ellis, George F. R.},
  title = {Republication of: Etherington's Reciprocity Theorem},
  journal = {General Relativity and Gravitation},
  volume = {39},
  pages = {1047--1052},
  year = {2007},
  eprint = {astro-ph/0703759},
  archivePrefix = {arXiv},
  doi = {10.1007/s10714-007-0445-5}
}

@article{Chevallier:2000qy,
  author = {Chevallier, M. and Polarski, D.},
  title = {Accelerating Universes with Scaling Dark Matter},
  journal = {International Journal of Modern Physics D},
  volume = {10},
  pages = {213--224},
  year = {2001},
  eprint = {gr-qc/0009008},
  archivePrefix = {arXiv},
  doi = {10.1142/S0218271801000822}
}

@article{Linder:2002et,
  author = {Linder, Eric V.},
  title = {Exploring the Expansion History of the Universe},
  journal = {Physical Review Letters},
  volume = {90},
  pages = {091301},
  year = {2003},
  eprint = {astro-ph/0208512},
  archivePrefix = {arXiv},
  doi = {10.1103/PhysRevLett.90.091301}
}

@article{Brout:2022vxf,
  author = {Brout, Dillon and others},
  title = {The Pantheon+ Analysis: Cosmological Constraints},
  journal = {Astrophysical Journal},
  volume = {938},
  pages = {110},
  year = {2022},
  eprint = {2202.04077},
  archivePrefix = {arXiv},
  primaryClass = {astro-ph.CO},
  doi = {10.3847/1538-4357/ac8e04}
}

@INPROCEEDINGS{2016AAS...22713918R,
       author = {{Rubin}, David and {Aldering}, Greg Scott and {Amanullah}, Rahman and {Barbary}, Kyle H. and {Bruce}, Adam and {Chappell}, Greta and {Currie}, Miles and {Dawson}, Kyle S. and {Deustua}, Susana E. and {Doi}, Mamoru and {Fakhouri}, Hannah and {Fruchter}, Andrew S. and {Gibbons}, Rachel A. and {Goobar}, Ariel and {Hsiao}, Eric and {Huang}, Xiaosheng and {Ihara}, Yutaka and {Kim}, Alex G. and {Knop}, Robert A. and {Kowalski}, Marek and {Krechmer}, Evan and {Lidman}, Chris and {Linder}, Eric and {Meyers}, Joshua and {Morokuma}, Tomoki and {Nordin}, Jakob and {Perlmutter}, Saul and {Ripoche}, Pascal and {Ruiz-Lapuente}, Pilar and {Rykoff}, Eli S. and {Saunders}, Clare and {Spadafora}, Anthony L. and {Suzuki}, Nao and {Takanashi}, Naohiro and {Yasuda}, Naoki and {Supernova Cosmology Project}},
        title = "{The Union3 Supernova Ia Compilation}",
    booktitle = {American Astronomical Society Meeting Abstracts \#227},
         year = 2016,
       series = {American Astronomical Society Meeting Abstracts},
       volume = {227},
        month = jan,
          eid = {139.18},
        pages = {139.18},
       adsurl = {https://ui.adsabs.harvard.edu/abs/2016AAS...22713918R}
}

@article{Torrado:2020xyz,
  author = {Torrado, Jesus and Lewis, Antony},
  title = {Cobaya: Code for Bayesian Analysis of hierarchical physical models},
  journal = {Journal of Cosmology and Astroparticle Physics},
  volume = {2021},
  number = {05},
  pages = {057},
  year = {2021},
  eprint = {2005.05290},
  archivePrefix = {arXiv},
  primaryClass = {astro-ph.IM},
  doi = {10.1088/1475-7516/2021/05/057}
}

@article{Shafieloo:2012ht,
  author = {Shafieloo, Arman},
  title = {Crossing Statistic: Reconstructing the Expansion History of the Universe},
  journal = {Journal of Cosmology and Astroparticle Physics},
  volume = {2012},
  number = {05},
  pages = {024},
  year = {2012},
  eprint = {1202.4808},
  archivePrefix = {arXiv},
  primaryClass = {astro-ph.CO},
  doi = {10.1088/1475-7516/2012/05/024}
}

@article{Sahni:2008xx,
    author = "Sahni, Varun and Shafieloo, Arman and Starobinsky, Alexei A.",
    title = "{Two new diagnostics of dark energy}",
    eprint = "0807.3548",
    archivePrefix = "arXiv",
    primaryClass = "astro-ph",
    doi = "10.1103/PhysRevD.78.103502",
    journal = "Phys. Rev. D",
    volume = "78",
    pages = "103502",
    year = "2008"
}

@article{Dinda:2025hiu,
    author = "Dinda, Bikash R. and Maartens, Roy and Clarkson, Chris",
    title = "{Calibration-independent consistency test of DESI DR2 BAO and SNIa}",
    eprint = "2509.19899",
    archivePrefix = "arXiv",
    primaryClass = "astro-ph.CO",
    doi = "10.1088/1475-7516/2025/12/025",
    journal = "JCAP",
    volume = "12",
    pages = "025",
    year = "2025"
}

@article{Dinda:2026uff,
    author = "Dinda, Bikash R. and Maartens, Roy and Clarkson, Chris",
    title = "{Calibration-independent consistency test of BAO and SNIa data: update}",
    eprint = "2601.16229",
    archivePrefix = "arXiv",
    primaryClass = "astro-ph.CO",
    month = "1",
    year = "2026"
}

@ARTICLE{2012JCAP...05..024S,
       author = {{Shafieloo}, Arman},
        title = "{Crossing statistic: Bayesian interpretation, model selection and resolving dark energy parametrization problem}",
      journal = {\jcap},
         year = 2012,
        month = may,
       volume = {2012},
       number = {5},
          eid = {024},
        pages = {024},
          doi = {10.1088/1475-7516/2012/05/024},
archivePrefix = {arXiv},
       eprint = {1202.4808},
 primaryClass = {astro-ph.CO},
       adsurl = {https://ui.adsabs.harvard.edu/abs/2012JCAP...05..024S}
}

@ARTICLE{2013JCAP...04..042S,
       author = {{Shafieloo}, Arman and {Majumdar}, Subhabrata and {Sahni}, Varun and {Starobinsky}, Alexei A.},
        title = "{Searching for systematics in SNIa and galaxy cluster data using the cosmic duality relation}",
      journal = {\jcap},
         year = 2013,
        month = apr,
       volume = {2013},
       number = {4},
          eid = {042},
        pages = {042},
          doi = {10.1088/1475-7516/2013/04/042},
archivePrefix = {arXiv},
       eprint = {1212.1277},
 primaryClass = {astro-ph.CO},
       adsurl = {https://ui.adsabs.harvard.edu/abs/2013JCAP...04..042S}
}

@article{Matthewson_2025,
   title={Star-crossed labours: checking consistency between current supernovae compilations},
   volume={2025},
   ISSN={1475-7516},
   url={http://dx.doi.org/10.1088/1475-7516/2025/01/064},
   DOI={10.1088/1475-7516/2025/01/064},
   number={01},
   journal={Journal of Cosmology and Astroparticle Physics},
   publisher={IOP Publishing},
   author={Matthewson, W.L. and Shafieloo, A.},
   year={2025},
   month=jan, pages={064} }

@article{Gonzalez-Fuentes:2025lei,
    author = "Gonz{\'a}lez-Fuentes, Alex and G{\'o}mez-Valent, Adri{\`a}",
    title = "{Reconstruction of dark energy and late-time cosmic expansion using the Weighted Function Regression method}",
    eprint = "2506.11758",
    archivePrefix = "arXiv",
    primaryClass = "astro-ph.CO",
    doi = "10.1088/1475-7516/2025/12/049",
    journal = "JCAP",
    volume = "12",
    pages = "049",
    year = "2025"
}

@article{DESI:2024aqx,
    author = "Calderon, R. and others",
    collaboration = "DESI",
    title = "{DESI 2024: reconstructing dark energy using crossing statistics with DESI DR1 BAO data}",
    eprint = "2405.04216",
    archivePrefix = "arXiv",
    primaryClass = "astro-ph.CO",
    doi = "10.1088/1475-7516/2024/10/048",
    journal = "JCAP",
    volume = "10",
    pages = "048",
    year = "2024"
}

@book{Press2007Numerical,
  author = {Press, William H. and Teukolsky, Saul A. and Vetterling, William T. and Flannery, Brian P.},
  day = 10,
  edition = 3,
  howpublished = {Hardcover},
  isbn = {0521880688},
  month = sep,
  publisher = {Cambridge University Press},
  title = {Numerical Recipes 3rd Edition: The Art of Scientific Computing},
  year = 2007
}

@article{Hwang:2022hla,
    author = "Hwang, Seung-gyu and L'Huillier, Benjamin and Keeley, Ryan E. and Jee, M. James and Shafieloo, Arman",
    title = "{How to use GP: effects of the mean function and hyperparameter selection on Gaussian process regression}",
    eprint = "2206.15081",
    archivePrefix = "arXiv",
    primaryClass = "astro-ph.CO",
    doi = "10.1088/1475-7516/2023/02/014",
    journal = "JCAP",
    volume = "02",
    pages = "014",
    year = "2023"
}

@article{DES:2024jxu,
    author = "Abbott, T. M. C. and others",
    collaboration = "DES",
    title = "{The Dark Energy Survey: Cosmology Results with {\ensuremath{\sim}}1500 New High-redshift Type Ia Supernovae Using the Full 5 yr Data Set}",
    eprint = "2401.02929",
    archivePrefix = "arXiv",
    primaryClass = "astro-ph.CO",
    reportNumber = "FERMILAB-PUB-23-0821-PPD, DES-2023-805",
    doi = "10.3847/2041-8213/ad6f9f",
    journal = "Astrophys. J. Lett.",
    volume = "973",
    number = "1",
    pages = "L14",
    year = "2024"
}

@article{DES:2025sig,
    author = "Popovic, B. and others",
    collaboration = "DES",
    title = "{The Dark Energy Survey Supernova Program: A Reanalysis Of Cosmology Results And Evidence For Evolving Dark Energy With An Updated Type Ia Supernova Calibration}",
    eprint = "2511.07517",
    archivePrefix = "arXiv",
    primaryClass = "astro-ph.CO",
    reportNumber = "FERMILAB-PUB-25-0842-CSAID-PPD",
    month = "11",
    year = "2025"
}

@ARTICLE{2010PhRvD..82j3502H,
       author = {{Holsclaw}, Tracy and {Alam}, Ujjaini and {Sans{\'o}}, Bruno and {Lee}, Herbert and {Heitmann}, Katrin and {Habib}, Salman and {Higdon}, David},
        title = "{Nonparametric reconstruction of the dark energy equation of state}",
      journal = {Physical Review D},
         year = 2010,
        month = nov,
       volume = {82},
       number = {10},
          eid = {103502},
        pages = {103502},
          doi = {10.1103/PhysRevD.82.103502},
archivePrefix = {arXiv},
       eprint = {1009.5443},
 primaryClass = {astro-ph.CO},
       adsurl = {https://ui.adsabs.harvard.edu/abs/2010PhRvD..82j3502H}
}

@ARTICLE{2012PhRvD..85l3530S,
       author = {{Shafieloo}, Arman and {Kim}, Alex G. and {Linder}, Eric V.},
        title = "{Gaussian process cosmography}",
      journal = {Physical Review D},
         year = 2012,
        month = jun,
       volume = {85},
       number = {12},
          eid = {123530},
        pages = {123530},
          doi = {10.1103/PhysRevD.85.123530},
archivePrefix = {arXiv},
       eprint = {1204.2272},
 primaryClass = {astro-ph.CO},
       adsurl = {https://ui.adsabs.harvard.edu/abs/2012PhRvD..85l3530S}
}

@ARTICLE{2008IJMPD..17.2315L,
       author = {{Linder}, Eric V. and {Miquel}, Ramon},
        title = "{Cosmological Model Selection:. Statistics and Physics}",
      journal = {International Journal of Modern Physics D},
         year = 2008,
        month = jan,
       volume = {17},
       number = {12},
        pages = {2315-2324},
          doi = {10.1142/S0218271808013881},
archivePrefix = {arXiv},
       eprint = {astro-ph/0702542},
 primaryClass = {astro-ph},
       adsurl = {https://ui.adsabs.harvard.edu/abs/2008IJMPD..17.2315L}
}

\end{document}